\documentclass[twocolumn,secnumarabic,amssymb, nobibnotes, aps, prd,nofootinbib, preprintnumbers]{revtex4-1}

\usepackage[T1]{fontenc}
\usepackage[utf8]{inputenc}
\usepackage[english]{babel}
\usepackage{amsmath}
\usepackage{amssymb}
\usepackage{tikz}		
\usepackage{adjustbox}		
\usepackage{enumitem}
\usepackage{mathtools}
\usepackage{booktabs}
\usepackage{multirow}
\usepackage{subfig}
\usepackage{dsfont}
\usepackage{feynmp-auto}
\usepackage{color}
 \usepackage{cancel}
\usepackage{hyperref}
\usepackage{lipsum}

\DeclarePairedDelimiter\abs{\lvert}{\rvert}%
\DeclarePairedDelimiter\norm{\lVert}{\rVert}%

\makeatletter
\let\oldabs\abs
\def\abs{\@ifstar{\oldabs}{\oldabs*}}
\let\oldnorm\norm
\def\norm{\@ifstar{\oldnorm}{\oldnorm*}}
\makeatother

\begin{document}
\title{Non-minimal $331$ model for Lepton Flavour Universality Violation in $b\to s\ell\ell$ decays}%
\preprint{LPT-Orsay-17-49}
\preprint{TTP17-044}

\author{S. Descotes-Genon}%
\affiliation{Laboratoire de Physique Th\'{e}orique, \\UMR 8627, CNRS, Univ. Paris-Sud, Universit\'{e} Paris-Saclay, 91405 Orsay Cedex, France}

\author{M. Moscati}%
\affiliation{Institute for Theoretical Particle Physics (TTP), Karlsruhe Institute of Technology, \\Engesserstra{\ss}e 7, D-76128 Karlsruhe, Germany}

\author{G. Ricciardi}%
\affiliation{Dipartimento di Fisica E. Pancini, Universit\`{a} di Napoli Federico II,  \\
Complesso Universitario di Monte Sant'Angelo,Via Cintia, 80126 Napoli, Italy}
\affiliation{I.N.F.N. Sezione di Napoli, \\
Complesso Universitario di Monte Sant'Angelo, Via Cintia, 80126 Napoli, Italy}

\email[]{sebastien.descotes-genon@th.u-psud.fr, marta.moscati@kit.edu, giulia.ricciardi@na.infn.it}

\begin{abstract}331 models constitute an extension of the Standard Model (SM) obtained by enlarging the SM gauge group $SU(3)_\text{C}\times SU(2)_\text{L}\times U(1)_Y$ to the group $SU(3)_\text{C}\times SU(3)_\text{L}\times U(1)_X$. We investigate how a non-minimal 331 model may embed lepton flavour universality violating  contributions to $b\to s\ell\ell$ processes without introducing lepton flavour violation, as suggested by the recent LHCb measurements of the ratios $R_K$ and $R_{K^*}$. We discuss the model-independent scenarios of New Physics in $b\to s\ell\ell$ currently favoured by the data that could be accommodated by this model and consider a few phenomenological constraints on this model.
\end{abstract}

\maketitle
\section{Introduction}
\label{sec:intro}
At the energies currently reached at the LHC, no direct signals of New Physics (NP) have arisen yet, in the sense that only particles already  in the Standard Model (SM) have been observed directly.
 This
has pushed the scale of many NP  models much above the electroweak scale, challenging the earlier expectations that these two scales would be similar for these models -- supersymmetric models being the most prominent ones.

On the other hand, recent disagreements with the SM expectations have appeared in flavour physics and more specifically in  $b$-quark decays (for recent reviews see Refs.~\cite{Ricciardi:2016jjb,Capdevila:2017ert, Koppenburg:2017mad, Ricciardi:2016pmh} and references therein).
In particular, four anomalies have appeared in ratios assessing Lepton Flavour Universality (LFU) in the decays $B\to K^{(*)} \ell^+ \ell^-$ (corresponding to the quark-level decay $b\to s\ell\ell$) and $B\to D^{(*)} \ell\bar\nu_\ell$ (corresponding to the quark-level decay $b\to c\ell\nu$), where $\ell$ stands for $e, \mu, \tau$. The ratios of current interest are defined as
\begin{equation}\begin{split}
	&R_{K^{(*)}[q^2_{\rm min},q^2_{\rm max}]}=\frac{\mathcal B(B\to K \mu^+\mu^-)_{q^2\in[q^2_{\rm min},q^2_{\rm max}]}}{\mathcal B(B\to K e^+e^-)_{q^2\in[q^2_{\rm min},q^2_{\rm max}]}}\\
	&R_{D^{(*)}}=\frac{\mathcal B(B\to D^{(*)} \tau\bar\nu_\tau)}{\mathcal B(B\to D^{(*)} \ell\bar\nu_\ell)},\quad [\ell=e, \mu]
\end{split}\end{equation}~\\
where $R_{K^{(*)}}$ are measured over specific ranges for the squared di-lepton invariant mass $q^2$ (in GeV$^2$), whereas $R_{D^{(*)}}$ deals with the total branching ratios.
It is interesting to make a comparison between the experimental and theoretical values for these quantities:
\begin{widetext}
\begin{equation}
\begin{aligned}[l]
	&R_{K[1,6]}^\text{exp}=0.745^{+0.090}_{-0.074}\pm 0.036\text{~\cite{Rk:LHCb}}\\
	&R_{K^*[0.045,1.1]}^\text{exp}=0.66^{+0.11}_{-0.07}\pm 0.03\text{~\cite{Rkst:LHCb}}\\
	&R_{K^*[1.1,6.0]}^\text{exp}=0.69^{+0.11}_{-0.07}\pm 0.05\text{~\cite{Rkst:LHCb}}\\
	&R_{D}^\text{exp}=0.407\pm 0.039 \pm 0.024\text{~\cite{RD:HFAG}}\\
	&R_{D^{*}}^\text{exp}=0.304\pm 0.013 \pm 0.007\text{~\cite{RD:HFAG}}
\end{aligned}
\quad
\begin{aligned}[l]
	&R_K^\text{th}=1.00\pm 0.01\text{~\cite{Bordone:2016gaq,Capdevila:2017ert}}\\
	&R_{K^*[0.045,1.1]}^\text{th}=0.922\pm 0.022\text{~\cite{Capdevila:2017ert}}\\
	&R_{K^*[1.1,6.0]}^\text{th}=1.000\pm 0.006\text{~\cite{Capdevila:2017ert}}\\
	&R_{D}^\text{th}=0.300 \pm 0.008\text{~\cite{RDs:th}}\\
	&R_{D^{*}}^\text{th}=0.252\pm0.003\text{~\cite{Fajfer:2012vx}}.
\end{aligned}
\quad
\begin{aligned}[l]
	& 2.8~ \sigma\\
        & 2.7~ \sigma\\
        & 3.0~ \sigma\\
        & 2.3~ \sigma\\
        & 3.4~ \sigma\\
\end{aligned}
\end{equation}\end{widetext}
In the experimental data the first errors are statistical and the second ones systematic. Prominent contributions to these ratio determinations have been given by  Babar, Belle and LHCb \cite{RD:BaBar1, RD:BaBar2, RD:LHCb, RD:Belle, RD:abd, Rk:LHCb, Rkst:LHCb}.
Although it is still not excluded that the previous disagreements might be accounted to statistical fluctuations of the data, or to a possible underestimate of the theoretical errors, an interesting aspect of these anomalies lies in the fact that they all seem to point in the direction of a possible Lepton Flavour Universality Violation (LFUV) in the interactions mediating the processes. Moreover, another LFU ratio has been measured recently, corresponding again to the quark decay $b\to c\ell\nu_\ell$~\cite{Aaij:2017tyk}:
\begin{equation}
 	R_{J/\psi}=\frac{\mathcal B(B_c\to J/\psi \tau\bar\nu_\tau)}{\mathcal B(B_c\to J/\psi \mu\bar\nu_\mu)},
\end{equation}
around 2$\sigma$ above the SM predictions.

For what concerns the $R_K$ and $R_{K^*}$ anomaly, the situation becomes even more intriguing for three reasons. First of all, the process is mediated by a Flavour Changing Neutral Current (FCNC). Since such a current cannot arise at tree level in the SM, the suppression due to the loop structure implies that the possible contribution of NP effects might arise in a significant way in this process. Furthermore, it was noticed in Ref.~\cite{Hiller:2003js} that in the ratios $R_{K^{(*)}}$ the hadronic uncertainties cancel
to a very large extent~\cite{Hiller:2014ula, sdg,sdg2, bec,Rk:eff, Rk:eff2, sim:eff, Bordone:2016gaq}~\footnote{The same cancellation does not occur for $R_{D^{(*)}}$ due to the presence of the heavy lepton $\tau$ in the final state.}, thus reducing substantially the uncertainty on the theoretical expectations.
Finally, these deviations concerning the branching ratios are only a part of the anomalies observed in $b\to s\mu\mu$ decays. Contrary to $b\to c\ell\nu$ transitions, there are many other observables that have been measured, especially concerning the angular distribution of the decay products in the decays $B\to K^*(\to K\pi)\mu\mu$ and $B_s\to\phi(\to KK)\mu\mu$, and some observables (the so-called $P_2$ and $P_5'$~\cite{Matias:2012xw,DescotesGenon:2012zf,Descotes-Genon:2013vna}) have featured deviations from SM expectations in addition to the LFUV ratios quoted above~\cite{Aaij:2015oid,Aaij:2015esa,Wehle:2016yoi,Abdesselam:2016llu,Aaij:2014pli}.
Many model-independent analyses of these anomalies in $b\to s\ell\ell$ have already been performed in terms of effective field theories corresponding to the SM at the $b$-quark mass scale, supplemented with the additional lowest dimensional non-SM operators~\cite{sdg, sdg2, bec, Rk:eff, Rk:eff2, sim:eff,Capdevila:2017bsm,Altmannshofer:2017yso, Geng:2017svp,Ciuchini:2017mik,Hurth:2017hxg,Alok:2017jaf,Alok:2017sui,Choudhury:2017qyt}. They are able to accommodate all the deviations observed in $b\to s\ell\ell$  in terms of a significant shift of the short-distance Wilson coefficient $C_9^\mu$, possibly together with shifts in other Wilson coefficients such as $C_{9'}^\mu$ or $C_{10}^\mu$. Remarkably, the same shift is needed to explain
the anomalies in the angular observables in $B\to K^*\mu\mu$ and the LFUV ratios of branching ratios $R_{K^{(*)}}$.

While model-independent analyses are powerful tools to understand the pattern of the anomalies in terms of NP contributions already  felt at low energies, they are not able to provide a dynamical explanation for these deviations. This requires us to choose
specific scenarios of physics Beyond the Standard Model (BSM) and try to see if  they allow for such anomalies.
Several models have been proposed to account for $R_{K^{(*)}}$ and $R_{D^{(*)}}$ simultaneously. Most of the successful candidates can be cast in two sets~\cite{Buttazzo:2017ixm}.
One set includes models that try to reproduce the presence of LFUV by assuming that the relevant processes are mediated by leptoquark particles~(see, e.g., Refs.~\cite{lq1,lq2,lq3,lq4,lq5,lq6,DiLuzio:2017vat,Calibbi:2017qbu}).
In the other set the process is mediated by heavy exotic gauge bosons, whose couplings depend on the generation~(see, e.g., Refs.~\cite{Crivellin:2015mga,Crivellin:2015lwa,Crivellin:2015era,Greljo:2015mma,Boucenna:2016wpr}).
In this article, we analyse a model falling in the latter category, and corresponding to a specific version of the so-called 331 models~\cite{Pisano:1991ee,Frampton:1992wt}.

331 models constitute one of the simplest extensions of the SM~\cite{Lee:1977qs,Lee:1977tx,Buccella:1977gx,Buccella:1978nc}. The gauge group is extended from the SM gauge group $SU(3)_\text{C}\times SU(2)_\text{L}\times U(1)_Y$ to the group $SU(3)_\text{C}\times SU(3)_\text{L}\times U(1)_X$. These models experience thus two stages of breaking: at a heavier scale $\Lambda_{NP}$, the extended group is broken down to the SM gauge group, for which electroweak symmetry breaking occurs at the lower scale $\Lambda_{EW}$. Phenomenologically, these models feature heavy gauge bosons ($W'$, $Z'$) as well as an extended Higgs sector triggering the two spontaneous breakdowns, leading to heavy scalar/pseudoscalar bosons ($H$, $A$), with electric charges depending on the implementation of the model.

In the most studied version~\cite{331:buras, 331:burasmu,331:burasZZ, 331:anomaly, 331:higgs, ponce:331, sher:331,Queiroz:2016gif,Alves:2016fqe}, one simply extends each $SU(2)_\text L$ doublet to one of the two fundamental representations of $SU(3)_\text L$, namely either $3$ or $\bar 3$, without introducing any additional family. Furthermore, this assumption is taken  together with the requirement of cancellation of chiral anomalies, that prescribes that the number of triplets is equal to the number of antitriplets. The three lepton families are then forced to belong to the same fundamental representation of the group, hence implying the family-independence of the couplings with gauge bosons. This in turn prevents any LFUV at the level of the gauge couplings to the leptons.

Another version of the 331 model, partially analysed in Refs.~\cite{331:CAB1, 331:CAB2}, extends the lepton sector by introducing two additional generations. With this assumption, one ends up with a lepton generation that transforms differently compared to the others, and hence presents different couplings with the gauge bosons; this situation suffices to guarantee the presence of LFUV. Two, rather than one, additional lepton generations, are required to preserve anomaly cancellation. We will focus on this version of the 331 model, and we will study if it can reproduce the anomalies observed in $b\to s\ell\ell$ processes under simple assumptions: LFUV is present and dominated by neutral gauge boson contributions, there is no significant Lepton Flavour Violation (LFV) of the form
$b\to s\ell_1\ell_2$, the model should not yield too large contributions to $B_s\bar{B}_s$ mixing. It turns out that the model is then able to reproduce scenarios with large contributions to $(C_9^\mu,C_{10}^\mu)$ in good agreement with global fit analyses of $b\to s\ell\ell$.

The paper is organised as follows: in section~\ref{sec:331} we review  the main features of our model, and justify our choices compared to the minimal 331 models more often studied in the literature. In section~\ref{sec:LFUV} we analyse the gauge boson-mediated contributions arising for the process $b\to s \ell \ell$, pointing out the arising of LFUV in the couplings. In section~\ref{sec:global} we compare these contributions with the global analyses performed in Refs.~\cite{sdg, sdg2}. In section~\ref{sec:mixing}, we examine other simple phenomenological constraints on the model
for the gauge boson contributions considered here, in particular $B_s\bar{B}_s$ mixing.
 In section~\ref{sec:outlook} we conclude and discuss further extensions of the model, for instance concerning LFUV in $R_{D^{(*)}}$. Finally, the appendices are devoted to various computations concerning the spectrum and couplings of our model.

\section{Features of the 331 model}
\label{sec:331}

Starting from the gauge group $SU(3)_\text{C}\times SU(3)_\text{L}\times U(1)_X$ (with gauge couplings $g_S,g,g_X$), the model will undergo two Spontaneous Symmetry Breakings (SSB). The first one occurs at an energy scale $\Lambda_\text{NP}$ and allows to recover the SM gauge group. The subsequent one, at energy scale $\Lambda_\text{EW}$, reproduces the Electroweak Symmetry Breaking (EWSB) of the SM. We assume that $\Lambda_\text{NP}\gg\Lambda_\text{EW}$, and introduce a parameter $\epsilon=\Lambda_\text{EW}/\Lambda_\text{NP}$ keeping track of the order of magnitude of the NP contributions of the model.

When enlarging the SM gauge group, embedding it into the broader $SU(3)_\text{C}\times SU(3)_\text{L}\times U(1)_X$ group,
there are a few general requirements to be obeyed:
\begin{itemize}
\item
the model should contain representations consistent with the SM quantum numbers and should have no anomalies, which sets powerful constraints on the choice of representations for the fermions~\cite{331:anomaly},
\item it should exhibit a Higgs sector able to trigger the two stages of spontaneous symmetry breaking (breaking down to the SM group and electroweak symmetry breaking) and to generate masses with a hierarchy in agreement with the observations (no light particles apart from the SM ones)~\cite{331:higgs}.
\end{itemize} For our particular purposes, we will also require that the lepton generations are not embedded equally into $SU(3)_\text{L}$ representation, in order to be able to generate LFUV at the level of the interactions.

\subsection{Choice of $\beta$}
\label{choicebeta}

We start by discussing the generators of the $SU(3)_\text{L}$ group and its connection with the SM gauge group.
Leaving aside the case of $SU(3)_\text{C}$, that presents no differences with respect to the
SM, the generators of the $SU(3)_\text{L}$ gauge group are  indicated with $\hat T^1 \cdots \hat T^8$.
Since the  generator  of the $U(1)_X$ group must commute
with the generators of $SU(3)_\text{L}$, it has to be be proportional to the identity in the space
referred to the representation of $SU(3)_\text{L}$.
The normalisation of the generators is $\mathrm{Tr}[\hat T^i \, \hat T^j]=\delta^{ij}/2 $,  and ${\mathds 1} = \mathrm{diag} (1, 1, 1)$  is the identity matrix.
We define the $U(1)_X$ generator
 as
$\hat T^9 = {\mathds 1}/\sqrt{6}$, since  this definition  implies  the same normalisation relation as
the other eight generators.

We can then identify the hypercharge operator ${\hat Y}$ in terms of the generators of the new gauge group, by
requiring that $\hat Y$ commutes with all the generators of $SU(2)_\text{L}$, which forces it to have only terms proportional to $\hat T^8$ and to the $U(1)_X$ generator.
Naming $X$ the quantum number associated with $U(1)_X$,
we define
\begin{equation}
	\frac{\hat Y}{2} =\beta \hat T^8+X{\mathds 1}
\label{hyper:ID}
\end{equation}
where  $\hat T^8 = 1/2 \, \hat{\lambda}^8= 1/(2\sqrt{3}) \, \mathrm{diag}(1, 1,-2)$. With $\hat{\lambda}^i$ we indicate the Gell-Mann matrices.
With this definition of the hypercharge, the electric charge operator reads
\begin{equation}\hat Q=  a \hat T^3 +  \frac{\hat{Y}}{2}= a \hat T^3+ \beta \hat T^8+X{\mathds 1} \end{equation}
where $\hat T^3 = 1/2 \, \hat{\lambda}^3= 1/2 \, \mathrm{diag}(1, -1,0)$.
The electric charge is defined in general as a linear combination of the diagonal
generators of the group, where
 the value of the proportionality constant $a$ and $\beta$ distinguishes  different 331 models.

In order to obtain isospin doublets which embed
$SU (2)_\text{L} \times U (1)_Y$ into $SU (3)_L \times U (1)_X$, we set $a=1$.
The way
in which the SM electroweak gauge group is embedded in $SU(3)_\text{L} \times U(1)_X$
is encoded in the
 parameter $\beta$, which controls the relation between the hypercharge and the $\hat T^8$ generator of $SU(3)_\text{L}$.
In order to restrict $\beta$ we could demand that no new particle introduced in the model
has exotic charges (i.e. different from the SM ones). Let us see how this condition operates when fermions belong to a triplet or an anti-triplet of $SU(3)_\text{L}$. After the first stage of symmetry breaking at the scale $\Lambda_\text{NP}$,
the $SU(3)_\text{L} \times U(1)_X$ representations of the fermions are broken down to $SU(2)_\text{L} \times U(1)_Y$
representations as follows
\begin{eqnarray}
\left({\bf{3}}, \mathrm{x}\right) &\rightarrow& \left({\bf{2}}, \frac{\beta}{\sqrt{3}} + 2\mathrm{x} \right)+ \left({\bf{1}}, -\frac{2\beta}{\sqrt{3}} + 2\mathrm{x}  \right) \nonumber \\
\left(\bar {\bf{3}}, \mathrm{x}\right) &\rightarrow& \left({\bf{2}}, -\frac{\beta}{\sqrt{3}} + 2\mathrm{x}   \right)+ \left({\bf{1}}, \frac{2\beta}{\sqrt{3}} + 2\mathrm{x}  \right)
\label{break1}
\end{eqnarray}
As just shown in Eq.~(\ref{break1}), both the triplet and the anti-triplet representation of $SU(3)_\text{L}$ are broken down to a doublet plus a singlet of $SU(2)_\text{L}$.  Let us consider the case of the quarks. We will choose to identify the first two components of the triplet (or anti-triplet) with the SM doublet: their charges acquire the SM values only by setting the $U(1)_Y$ hypercharges to the SM values, that is
$\pm \beta/\sqrt{3}+ 2\mathrm{x}=1/3$. The last entry of the triplet (or antitriplet) will be an additional, massive, fermion (called ``exotic'' in the following), with an electric charge $ \mp \sqrt 3/2 \, \beta + 1/6$, that becomes either 2/3 or -1/3 only if we choose $\beta= \mp 1/\sqrt3$~\footnote{Let us recall that other common values chosen in the literature, $\beta= \pm \sqrt3$, while maintaining the SM charge for the $SU(2)_\text{L}$ doublet, introduce exotic  electric charges for the  $SU(2)_\text{L}$ singlets (5/3 and -4/3).}. One can easily check that the same discussion also holds in the case of the leptons, with a similar outcome~\cite{331:buras}.

In this work, we will pick the particular value \begin{equation} \beta= -1/\sqrt3 \end{equation} It  can be related to the choice $ \beta=1/\sqrt{3} $ by changing all
the representations for their conjugates, and taking the opposite sign for the $U_X(1)$ charges. We will thus have the following definition of the electric charge operator
\begin{equation}\hat Q=  \hat T^3-\frac{1}{\sqrt3} \hat T^8+X{\mathds 1}
\label{charge1}
\end{equation}

\subsection{Fields and representations}\label{sec:fieldsandrepr}

In the following, we label the SM fermions with lower cases and the exotic ones with upper cases, choosing letters recalling their electric charge assignments. Using the notation $(SU(3)_\text C, SU(3)_\text L, U_X(1))$ while referring to the representations of the particles, we introduce the following fermionic content, which ensures the cancellation of the anomalies but allows for different representations for the three lepton generations, and thus potential LFUV effects (see also App.~\ref{app:fermioniccontent} for a summary of the representations chosen).

For the left-handed components, we introduce~\cite{ponce:331,sher:331,331:CAB1, 331:CAB2}
\begin{itemize}
\item three generations of quarks
\begin {equation}
\begin{split}
Q^L_m&=\begin{pmatrix}d^L_m\\-u^L_m\\B^L_m\end{pmatrix}\sim (3, \bar 3, 0), \quad m=1,2 \\
Q^L_3&=\begin{pmatrix}u^L_3\\d^L_3\\T^L_3\end{pmatrix}\sim (3, 3, \frac 1 3);
\end{split}\end{equation}
\item five generations of leptons
\begin {equation}
\begin{split}
\ell^L_1&=\begin{pmatrix}e^{-L}_1 \\ -\nu^L_1 \\ E^{-L}_1\end{pmatrix}\sim (1, \bar3, -\frac 2 3), \\
\ell^L_n&=\begin{pmatrix}\nu^L_n\\ e^{-L}_n \\N^{0L}_n\end{pmatrix}\sim (1, 3, -\frac 1 3), \qquad n=2,3 \\
\ell^L_4&=\begin{pmatrix}N^{0L}_4\\ E^{-L}_4 \\ P^{0L}_4\end{pmatrix}\sim (1, 3, -\frac 1 3),  \\
\ell^L_5&=\begin{pmatrix}\bigl(E^{-R}_4\bigr)^c\\ N^{0L}_5 \\ \bigl(e^{-R}_3\bigr)^c\end{pmatrix}\sim (1, 3, \frac 2 3). \\
\end{split}
\label{lh}
\end{equation}
\end{itemize}
The superscripts refer to the charge and the chirality of the fields. No  positively charged leptons have been introduced in the triplets. Indeed, they would  only appear in $\ell^L_5$, but  we identify them with the charge conjugate of the right-handed component of $E^{-}_4$ and $e^{-}_3$.
This identification avoids the presence of charged exotic particles with masses  of the order of the electroweak scale, which have not been observed~\footnote{We discuss the structure of the fermion masses derived from the Yukawa interactions between scalar and fermions in App.~\ref{app:higgs}, and in particular the masses of the charged leptons in App.~\ref{app:fermionmasses1}.}.

For the  right-handed components, we do not consider right handed partners for neutral particles, since they would be pure singlets with respect to the whole gauge group and of no relevance in our analysis (they should be added to discuss the neutrino mass matrix, which is beyond the scope of this article).
We define
\begin{itemize}
\item the quark fields
\begin {equation}
\begin{split}
d^R_{1,2,3}&\sim(3,1,-1/3)\\ B^R_m&\sim(3,1,-1/3),\qquad m=1,2\\
u^R_{1,2,3}&\sim(3,1,2/3)\\ T^R_3&\sim(3,1,2/3)
\end{split}
\end{equation}
\item the charged lepton fields
\begin {equation}
\begin{split}
&e^{-R}_{1,2}\sim(1,1,-1),\qquad E^{-R}_{1}\sim(1,1,-1)
\end{split}
\label{rh}
\end{equation}
\end{itemize}
As already indicated, the right-handed parts of $e_3^-$ and $E_4^-$ are not singlets, but belong to the lepton triplet $\ell^L_5$.

This particle content enables the cancellation of chiral anomalies. For instance, as discussed in Sec.~\ref{sec:intro}, it is easy to see that  the number of left-handed fermion triplets is equal to the number of left-handed fermion anti-triplets (taking into account that the quark fields are counted three times more than the lepton ones due to colour). Minimal 331 models  also exhibit the anomaly cancellation by having different $SU(3)_\textrm{L}$ representations for the three quark generations, but having the same representation for the three lepton generations prevents these minimal models from exhibiting LFUV. More details on the requirements imposed by the cancellation of anomalies can be found in App.~\ref{app:anomalycancel}.

It proves easier to discuss the spectrum of the theory after introducing the flavour vectors gathering fields with the same electric charge (for simplicity, we leave out the neutrino fields)
\begin{equation}\begin{split}
	D=\begin{pmatrix} d_1&d_2&d_3&B_1&B_2\end{pmatrix}^T,\qquad U=&\begin{pmatrix}u_1&u_2&u_3&T_3\end{pmatrix}^T,\\ 	f^-=\begin{pmatrix}
		e^-_1&
		e^-_2&
		e^-_3&
		E^-_1&
		E^-_4
	\end{pmatrix}&^T.
	\label{eq:flavvec}\end{split}
\end{equation}
We also group the $SU(3)_\text L$ gauge bosons as
\begin{equation}\begin{split}
	W_\mu&=W_\mu^aT^a=\\&=\frac 1 2 \begin{pmatrix}
			W_\mu^3+\frac 1 {\sqrt 3} W_\mu^8&\sqrt 2 W_\mu^+&  W^4_\mu- i W_\mu^5\\
			\sqrt 2 W_\mu^-& -W_\mu^3+\frac 1 {\sqrt 3} W_\mu^8& W^6_\mu- i W_\mu^7\\
			W^4_\mu+ i W_\mu^5&W^6_\mu+ i W_\mu^7 & -\frac 2 {\sqrt 3} W_\mu^8
				\end{pmatrix}
\end{split}\end{equation}~\\~
and introduce
\begin{equation}\begin{split}
W^{\pm}_\mu =\frac 1 {\sqrt 2} (W^1_\mu\mp i W_\mu^2) &\qquad
V^{\pm}_\mu =\frac 1 {\sqrt 2} (W^6_\mu\mp i W_\mu^7) \\
Y^{0( 0 \star)}_\mu &=\frac 1 {\sqrt 2} (W^4_\mu\mp i W_\mu^5)\end{split}
\end{equation}
The values of the charges of the $V_\mu$ and $Y_\mu$ bosons  depend  on the value of $\beta$ (indeed, in the case $\beta=1/\sqrt 3$, we would have $V^{0( 0 \star)}_\mu$ and $Y^{\pm}_\mu$). Let us observe that for  $\beta=1/\sqrt 3$, $W^{4,5}$ are both eigenstates
of the charge operator with 0 eigenvalue, which allows the choice to use them, rather than $Y^{0( 0 \star)}$
as independent degrees of freedom.
We gather the interactions between the gauge bosons and the charged fermions in App.~\ref{app:curr}.

Summarising, we have chosen the particle content of the model in a way that allows LFUV, but otherwise  departs from the SM as little as possible. Fixing $\beta=-1/\sqrt{3}$ ensures  non-exotic charges for both SM and new fields in the spectra. Accommodating left-handed quarks  and left-handed leptons
in triplets or anti-triplets of $SU(3)_\text{L}$ representations, while assuming anomaly cancellation and LFUV simultaneously,  forces an unequal number of
 quark families and lepton families. We have allowed  the new degrees of freedom to be completely general, exception done for an identification in the fifth lepton generation and  the exclusion of  right-handed partners for neutral particles,  as justified above.
 This last assumption implies that no Dirac mass terms can be built for neutral particles (i.e., neutrinos).

\subsection{Symmetry breakings and spectrum}
\label{sec:SSBspec}

We are now in a position to discuss the two stages of symmetry breaking which will be assumed to be triggered by ($SU(3)_\text{C}$ singlet) scalar fields acquiring non-vanishing vacuum expectation values, in a way
analogous to the SM. On the other hand, we remain as general as possible for the representations under $SU(3)_\text{L}$, thus allowing for several scalar fields with different representations. The overall pattern of SSB is the following
\begin{widetext}
\begin{center}
\begin{adjustbox}{max width=\textwidth}
\begin{tikzpicture}
\node at (0,0) {$SU(3)_\text{C}\times SU(3)_\text{L} \times U(1)_X$};
\draw [->] (2.5,0) --  node[above] {$\chi,S_1$}node[below] {$\Lambda_\text{NP}$}  (3.5,0);
\node at (6,0) {$SU(3)_\text{C}\times SU(2)_\text{L} \times U(1)_Y$};
\draw [->] (8.5,0) --  node[above] {$\eta, \rho, S_{b,c}$}node[below] {$\Lambda_\text{EW}$}  (9.5,0);
\node at (10.5,0) {$U(1)_\text{EM}$};
\end{tikzpicture}
\end{adjustbox}
\end{center}\end{widetext}
The $SU(3)_\text{L}$ symmetry breaking  is accomplished through a triplet $\chi$ and a sextet  $S_1$.
The EWSB is accomplished by means of two triplets $\eta, \rho$ and two sextets $S_{b,c}$. Details on the structure of the vacuum expectation values of these fields and on their quantum numbers can be found in App.~\ref{app:higgs}.

There are five gauge fields that acquire a mass
of the order of $\Lambda_\text{NP}$, whereas the three remaining gauge fields will become massive at the electroweak scale.
At the first SSB, the  neutral and charged gauge bosons,  $W^{4,5}$ and $V^{\pm}$, acquire a mass, whereas the two neutral gauge bosons $X, W^8$ yield a massive neutral gauge boson $Z'$ and a massless one $B$, with a mixing angle $\theta_{331}$:
\begin{equation}
\begin{pmatrix}Z'\\B\end{pmatrix}=\begin{pmatrix}\cos\theta_{331}&-\sin\theta_{331}\\\sin\theta_{331}&\cos\theta_{331}\end{pmatrix}\begin{pmatrix}X\\W^8\end{pmatrix},
\end{equation}
The angle $\theta_{331}$ can be found by singling out the $Z'$ field in the
sector of the Lagrangian including the masses of the gauge bosons, which stems from the covariant derivative in the Higgs
Lagrangian.
It yields
\begin{equation}
\sin\theta_{331}=\frac{g}{\sqrt{g^2+\frac{g_X^2}{18}}}\,,\qquad
\cos\theta_{331}=-\frac{\frac{g_X}{3\sqrt2}}{\sqrt{g^2+\frac{g_X^2}{18}}}.
\label{mixing:angle}
\end{equation}
At the first stage of SSB, the mixing among neutral gauge bosons only involves $X$ and $W^8$, but not $W^{4,5}$ since these two classes of fields do not show the same EW quantum numbers, which correspond then to the unbroken part of the group. This can be seen for instance acting on them with the generator $T_3$. After the EWSB, only the neutral gauge boson identified with the photon remains massless, consisting of an admixture of $B$ and $W^3$ described by the weak angle $\theta_W$. The two mixing angles obey the relation~\cite{331:buras}
\begin{equation}
	\tan\theta_W=-\sqrt 3 \cos \theta_{331}\,,
\qquad
g=-\frac{g_X \tan \theta_{331}}{3 \sqrt{2}}\,.
\label{eq:wein}
\end{equation}
 This is actually a very general feature of the 331 model, which can be written as $\cos \theta_{331} = \beta \tan\theta_W$, with a deep relation with the pattern of EWSB (see for instance Eq.~(2.28) in Ref.~\cite{Correia:2017vxa} where the mixing angle is shifted by $90^\circ$  with respect to our notation). In particular, it is  possible to write~\cite{331:buras}
 \begin{equation}
 \frac{g_X^2}{g^2}=\frac{6\sin^2\theta_W}{1-(1+\beta^2)\sin^2\theta_W}\,.
 \end{equation}
 As $\sin^2\theta_W$ is close to 0.25, the perturbativity condition imposes significant constraints on the range of validity
 of the 331 models in the case of $\beta=\pm\sqrt{3}$: the $SU(3)_\text{L}$ symmetry breaking must occur at most
 at a few TeV~\cite{Ng:1992st}. This problem of perturbativity does not affect our case $\beta=-1/\sqrt{3}$, allowing our model to have room for a higher scale of $SU(3)_\text{L}$ symmetry breaking and significantly heavier gauge bosons, and
 providing a good justification to expansions in $\epsilon=\Lambda_\text{EW}/\Lambda_\text{NP}$.

While the photon consists of an admixture of the $W^3$  and $B$ fields only, the neutral gauge boson $Z$ that acquires mass from EWSB includes additional components from the $Z'$ and $W^4$ fields. Nevertheless, the
diagonalisation of the neutral gauge boson mass matrix after both stages of symmetry breaking shows that
the components along the exotic fields are suppressed by $\epsilon^2$ or higher. We will see in the following that the $Z$ contribution to $b\to s \ell \ell$ involves a $b\to s$ transition already suppressed by $\epsilon^2$, and we will neglect
the additionally $\epsilon^2$-suppressed contributions to the transition coming from the $Z'$ and $W^4$ components of the $Z$ mass eigenstate (which we will treat as consisting only of $W^3$  and $B$ at this order).

The most general Yukawa Lagrangian that can be built with the scalar fields provides a (heavy) mass to all the exotic particles after the $SU(3)_\text{L}$ SSB, in agreement with phenomenological expectations. The mass matrices arising for the charged fermions after the two SSBs are discussed in Apps.~\ref{app:fermionmasses} and~\ref{app:fermionmasses1}. Performing a singular value decomposition of the
up-type and down-type mass matrices yields the definition of the unitary rotation matrices relating (unprimed) interaction eigenstates and (primed) mass eigenstates
\begin{equation}\begin{split}\label{eq:VWrotations}
D^L&=V^{(d)} D'^L\,, \qquad U^L=V^{(u)} U'^L\,,\\
D^R&=W^{(d)} D'^R\,, \qquad U^R=W^{(u)} U'^R\,.\end{split}
\end{equation}
Due to the presence of the exotic fermions, these flavour matrices are $4\times 4$ (for up-type quarks) or $5\times 5$ (for down-type quarks) unitary matrices.
If we perform this diagonalisation order by order in $\epsilon$, we observe the following pattern for the mixing matrices $V^{(u,d)}$ and $W^{(u,d)}$
\begin{itemize}
	\item at order $\epsilon^0$, the SM fields are massless and they only mix among themselves; the massive exotic particles  mix also only among themselves;
	\item at order $\epsilon^1$, there is only mixing between SM and exotic particles;
	\item the $\epsilon^2$ correction yields a mixing among all the particles of the same flavour vector.
\end{itemize}
This particular structure can be understood by diagonalising the mass matrix using perturbation theory in powers of $\epsilon$. Since the mass matrix for the SM particles is zero at $O(\epsilon^0)$, all SM particles are massless and degenerate at this order and they mix among themselves, whereas (heavy) exotic particles also mix among themselves. The normalisation of the eigenvectors require on the other hand that the $O(\epsilon^1)$ correction to an eigenvector is orthogonal to its $O(\epsilon^0)$ expression, leading to a $O(\epsilon^1)$ correction to the rotation matrix that mixes SM and exotic fields (but not SM fields alone, or exotic fields alone).

A remark is in order regarding the structure of the CKM matrix. This is given by the $W^+$ coupling with quarks,
which can be written as (see Eq.~\eqref{eq:W})
\begin{equation}\begin{split}
	\frac g {\sqrt2}W^+_\mu \bar U^L  \gamma^\mu \mathcal V D^L&=\frac g {\sqrt2}W^+_\mu\bar U^L\gamma^\mu \begin{pmatrix}1&0&0&0&0\\0&1&0&0&0\\0&0&1&0&0\\0&0&0&0&0\end{pmatrix}D^L\\
	&=V_{mn}^{CKM}\frac g {\sqrt2}W^+_\mu\bar{U}'^L_m \gamma^\mu D'^L_n
\end{split}\end{equation}
with the $4\times 5$ equivalent of the CKM matrix
\begin{equation}
	V^{CKM}=V^{(u)\dagger}\mathcal V V^{(d)}.
\end{equation}
Despite $V^{(u,d)}$ being unitary, the presence of $\mathcal V$ yields a non-unitary $V^{CKM}$ in the 331 model.
If we want to adequately reproduce the SM, we should however recover a unitary CKM matrix if we remain at low energies (i.e. leading order in $\epsilon$) and consider only the flavour subspace of the SM particles. As indicated above, at this order, the diagonalisation of the fermion mass terms occurs in a block-diagonal way: the mixing matrices $V^{(u)}$ and $V^{(d)}$ consist in two unitary blocks, one mixing the SM particles among themselves, and the other one mixing the exotic ones among themselves. Furthermore, $\mathcal V$ reduces to $\mathds{1}_{3\times3}$ in the SM flavour subspace. Therefore, at leading order in $\epsilon$, the $3\times 3$ SM block of $V^{CKM}$ will stem from the product of the two unitary $3\times 3$ SM subspaces of $V^{(u)}$ and $V^{(d)}$, ensuring that it is unitary at this order (this obviously does not mean that $V^{CKM}$ remains unitary at all orders in $\epsilon$, and this 331 model does indeed generate small deviations of unitarity for $V^{CKM}$).

A similar discussion could be held in the lepton sector, with the singular value decomposition of the charged lepton mass matrix leading to the definition of $5\times 5$ unitary rotation matrices between interaction and mass eigenstates
\begin{equation}
E^L=V^{(e)} E'^L\,, \qquad E^R=W^{(e)} E'^R\,.
\end{equation}
The PMNS matrix can  be built by combining unitary rotation matrices $V^{(e)}$ and $V^{(\nu)}$. A discussion of the PMNS matrix  would require a discussion of the neutrino spectrum, which is outside the scope of the present article.

\section{NP contribution to $b\to s \ell \ell$}
\label{sec:LFUV}

\subsection{Setting the problem}
\label{statementoftheproblem}

Having introduced a non-minimal 331 model with a SSB pattern leading to a phenomenologically viable spectrum, we will now investigate the consequences of the different representations for the lepton fields for LFUV in $b\to s\ell\ell$. We want to determine if this model is able to reproduce the pattern of deviations indicated in the current global analyses of this rare decay~\cite{Capdevila:2017bsm, Altmannshofer:2017yso, Geng:2017svp, Ciuchini:2017mik, Hurth:2017hxg}.

These analyses are performed in the framework of the effective Hamiltonian at the $b$-mass scale, separating short- and long-distance physics between Wilson coefficients and local operators~\cite{Grinstein:1987vj, Buchalla:1995vs}:
\begin{equation}
  {\mathcal H}_{\rm eff}=-\frac{4G_F}{\sqrt{2}} V_{tb} V_{ts}^* \sum_i C_i O_i
\end{equation}
 The main operators of interest for this discussion are the following:
\begin{equation}
\begin{split}
        O_7=&\frac{e}{16\pi^2} m_b (\bar\sigma_{\mu\nu} P_R b)F^{\mu\nu}\\
        O_{7'}=&\frac{e}{16\pi^2} m_b (\bar\sigma_{\mu\nu} P_L b)F^{\mu\nu}\\
	O_9^\ell=&\frac{e^2}{16\pi^2}(\bar s \gamma_{\mu} P_L b) (\bar\ell\gamma^\mu \ell)\\
	O_{10}^\ell=&\frac{e^2}{16\pi^2}(\bar s \gamma_{\mu} P_L b) (\bar\ell\gamma^\mu\gamma^5 \ell)\\
	O_{9'}^\ell=&\frac{e^2}{16\pi^2}(\bar s \gamma_{\mu} P_R b) (\bar\ell\gamma^\mu \ell)\\
	O_{10'}^\ell=&\frac{e^2}{16\pi^2}(\bar s \gamma_{\mu} P_R b) (\bar\ell\gamma^\mu\gamma^5 \ell).\\
\end{split}
\label{eq:OP}
\end{equation}
where $P_{L,R}=(1\mp \gamma_5)/2$ and the fields are understood as mass eigenstates. In the SM, only  $O_7$, $O_9^\ell$ and $O_{10}^\ell$ are significant, with values of the Wilson coefficients  $C_9^\ell\simeq 4.1$ and
$C_{10}^\ell\simeq -4.3$ at the scale $\mu=m_b$, whereas the  primed operators are $m_s/m_b$ suppressed due to the chirality of the quarks involved.

The analyses of the  $b\to s\gamma$ and $b\to s\ell\ell$ observables (both LFUV observables and angular observables for $b\to s\mu\mu$ and $b\to s\ell\ell$) point towards the fact that the pattern of deviations observed is consistent with  a large NP short-distance contribution to
$C_9^\mu$ (around 1/4 of the SM contribution)~\cite{sdg, sdg2,Capdevila:2017bsm}. More generally, scenarios with NP contributions in $C_9^\mu$ only, in $(C_9^\mu,C_{10}^\mu)$ or  in $(C_9^\mu,C_{9'}^\mu)$ are particularly favoured. On the other hand, the LFUV observables agree well with the absence of significant NP contributions to any electronic Wilson coefficients $C_{i}^{e}$.

 For the other operators, a good agreement with the SM is obtained: in other words, the fitted values of the NP contributions are constrained to remain small and these additional operators are not needed to improve the accuracy of the fit to the data. This is true for the operators suppressed in the SM, in particular scalar and pseudoscalar operators, which are constrained especially by the good agreement between the observed value for the $B_s\to\mu\mu$ branching ratio and its SM prediction. The same holds for the $O_7$ and $O_{7'}$ operators, which are constrained in particular by the $B\to X_s\gamma$ branching ratio.

\subsection{Gauge boson contributions}
\label{sec:gaugebosoncontributions}

 In view of these elements, we will focus on the vector/axial contributions which will be assumed to be the larger ones.
In particular, we will assume that the complex pattern of EWSB of our 331 model in the scalar potential ensures that the scalar/pseudoscalar contributions to $b\to s\ell\ell$ are small. This would correspond to constraints on the couplings $Y^{d},y^{d},j^{d},Y^{(-)},f^{(-)},y^{(-)},J,j,K,k,c$, the rotation matrices $V^{(d,e)},W^{(d,e)}$ and the masses of the six heavy scalar fields. In a similar way, we assume that the total NP contribution to $b\to s\gamma$ is small: as there are no $b\to s\gamma$ transitions at tree level in our model, the NP contribution would correspond to a sum of loops contributions involving a quark and either a neutral or a charged gauge boson or a heavy scalar bosons, i.e., involving the previous couplings, but also $Y^{u},y^{u},j^{u}$ and the rotation matrices $V^{(u)},W^{(u)}$. Let us mention that in both cases, the structure of the rotation matrices $V,W$ and the presence of the heavy masses ensure already that these NP contributions are somewhat suppressed. We could work out the parameter space of couplings, mixing and masses allowed by both types of constraints  in more detail, but at this stage, we are more interested in checking the constraints on the vector/axial sector, which are simpler and related to the deviations seen in $b\to s\ell\ell$ transitions.

The vector/axial contributions can only come from the neutral gauge bosons $Z', Z, A, W^{4,5}$. We will consider contributions at the lowest order in $\epsilon$ only, and we will focus
only on the non SM contribution to the Wilson coefficients (in other words, from now on $C_i=C_i^{NP}$).

Let us start with the interaction of $Z^\prime$ and $Z$ with the right-handed quarks. These interactions   are proportional to the identity in flavour space (see Eqs. \eqref{eq:Z'} and \eqref{eq:Z}), so
no flavour change can arise, at any order in $\epsilon$. We conclude that $Z^\prime$ and $Z$, do not contribute to $C'_{9,10}$ in the process $b\to s \ell^+\ell^-$. Only  contributions to $C_{9,10}$ are possible.

In the case of the heavy gauge boson $Z'$, a $O (\epsilon^2)$ suppression compared to the SM contribution comes from the heavy mass in the propagator of the gauge boson. The restriction of the interaction matrix to the SM particles is not proportional to the identity matrix in the interaction eigenbasis, as it can be seen in Sec.~\ref{sec:ZZprimecouplings}. Therefore, the flavour-changing transition $b\to s$  mediated by $Z'$  arises already after reexpressing
the interaction in the mass eigenbasis using the leading order $ \epsilon^0 $ rotation matrix. The overall suppression of the $Z'$ contribution is thus $O(\epsilon^2)$.
 Following Sec.~\ref{sec:ZZprimecouplings}, reexpressing the flavour eigenstates in the multiplets Eqs.~\eqref{eq:flavvec} in terms of mass eigenstates, and eliminating the coupling $g$ by means of  Eq.~\eqref{eq:wein}, we can rewrite the leading-order $Z'$ contribution in terms of effective operators as
{%
\begin{widetext}\begin{eqnarray}\label{eff:Z'}
	\mathcal H_{\text{eff}}&\supset &\frac{g_X^2}{54\cos^2\theta_{331}}\frac{1}{M^2_{Z'}}V^{(d)*}_{3k}V^{(d)}_{3l}
	\frac{4\pi}{\alpha}
	\\\nonumber
	&&	
	\Biggl\{\left[-\frac 1 2 V^{(e)*}_{1i}V^{(e)}_{1j}+\frac{1-6\cos^2\theta_{331}}2W^{(e)*}_{3i}W^{(e)}_{3j}+\frac{1+3\cos^2\theta_{331}}{4}\delta_{ij}\right]O^{klij}_9+\\
	&&\qquad+\left[\frac 1 2 V^{(e)*}_{1i}V^{(e)}_{1j}+\frac{1-6\cos^2\theta_{331}}2W^{(e)*}_{3i}W^{(e)}_{3j}+\frac{-1+9\cos^2\theta_{331}}{4}\delta_{ij}\right]O^{klij}_{10}\Biggr\}.\nonumber
\end{eqnarray}\end{widetext}
where the indices $k,l$ refer to the SM generations of the quark mass eigenstates (assuming $k\neq l$),  while $i, j$ refer to the SM lepton mass eigenstates
(either from the same or different generations).
The effective operators $O_{{9,10}}^{klij}$ are defined exactly as in Eq.~\eqref{eq:OP}, corresponding to the
$(\bar q_k\, q_l) (\bar \ell_i\, \ell_j)$ flavour structure. The fine-structure constant is $\alpha=e^2/(4 \pi)$. The $V$ and $W$ matrices provide the mixing matrices arising from the diagonalisation of the EWSB mass terms in the subspace of left-handed and right-handed SM fields. We stress that these rotations are related but cannot be identified with  the CKM or PMNS matrices and
 they can be considered only at order $\epsilon^0$ for our purposes (we have exploited their unitarity at that order for the $\delta_{ij}$ contributions).
 We notice that the presence of the mixing matrices yields LFUV couplings, and moreover
 a leptonic $i\neq j$ contribution might arise, corresponding to lepton-flavour violating transitions $b\to s \ell^+\ell'^-$, with different leptons in the final state, $\ell\neq\ell'$, which is a frequent feature of models generating LFUV couplings~\cite{Glashow:2014iga}.

 We can follow the same lines as the general analysis of the NP corrections to the effective Hamiltonian induced by neutral currents presented in App.~\ref{sec:ZZprimecouplings} and specialised to the case where the quarks have different flavours.

In the case of the SM gauge boson $Z$, there is no $b\to s$ transition allowed at order $\epsilon^0$, since the $3\times 3$ unitary rotation matrices restricted to the SM subspace cancel, following the same arguments as the discussion of the unitarity of the CKM matrix  at the end of Sec.~\ref{sec:SSBspec}. The transition does not arise at order $\epsilon^1$ either, since there is no correction to the mixing between SM particles at this order. The mixing between SM particles, leading to potential FCNC currents, starts only
at order $O(\epsilon^2)$. Since there is no suppression due to the mass of the intermediate gauge boson here, we conclude that the NP contribution from the SM gauge boson $Z$ starts at $O(\epsilon^2)$,  the same order as the $Z'$ contribution, although for different reasons. Indeed, starting from the interaction eigenbasis and switching  to the  mass eigenstates, we can express the part of interaction relevant to the process as
\begin{equation}\begin{split}
	\mathcal L_Z \supset  g \cos\theta_W Z_\mu\Bigl\{&\frac {1+3\cos^2\theta_{331}}2 \sum_\lambda \hat{V}^{(d)*}_{\lambda k}\hat{V}^{(d)}_{\lambda l} 
\bar D'^L_k\gamma^\mu D'^L_l\\
+&\frac{-1+3\cos^2\theta_{331}}2\bar f'^{-L}\gamma^\mu f'^{-L}+ \\	
+&3\cos^2\theta_{331}\bar f'^{-R}\gamma^\mu f'^{-R}\Bigr\},
\label{eq:Zcontr1}\end{split}
\end{equation}
 $\hat{V}^{(d)}$ represents the $O(\epsilon^1)$ correction to the rotation matrix $V^{(d)}$ between interaction and mass eigenstates for the left-handed down sector. As stated earlier, $\hat{V}^{(d)}_{mn}=0$ if $m$ and $n$ are both SM or both exotic, which means that the sum over $\lambda$ is restricted to exotic components here (as $k,l$ are SM components).
 Since the NP quark coupling to the $Z$ gauge boson is already of order $O(\epsilon^2)$, we need only the $O(\epsilon^0)$ coupling to the charged leptons. Due to the unitary block structure of the mixing matrix at this order and the structure of the $Z$ coupling to SM leptons (proportional to identify), we see that the rotation matrices cancel out, leading to the diagonal structure
 indicated in the leptonic sector of Eq.~\eqref{eq:Zcontr1}. In terms of effective operators and adopting the same notation of Eq.~\eqref{eff:Z'}, Eq.~\eqref{eq:Zcontr1} can be rewritten as
\begin{equation}\begin{split}
	\mathcal H_{\text{eff}}\supset \frac{\cos^2\theta_W(1+3\cos^2\theta_{331})}{8}\frac{g^2}{M^2_{Z}}\frac{4\pi}{\alpha}\sum_\lambda \hat{V}^{(d)*}_{\lambda k}\hat{V}^{(d)}_{\lambda l}\delta_{ij}\times\\\times\Bigl\{(-1+9\cos^2\theta_{331})O^{klij}_9+(1+3\cos^2\theta_{331})O^{klij}_{10}\Bigr\}.
\label{eff:Z}\end{split}
\end{equation}
We observe that the coupling is the same for all the light leptons, i.e. non-universality does not arise at order $\epsilon^2$ in the interaction with $Z$.
By comparing  Eq.~\eqref{eff:Z'} and Eq.~\eqref{eff:Z}, we explicitly see that although the non standard coupling originated from the $Z$ boson are  suppressed of order $\epsilon^2$ with respect to the ones of the $Z^\prime$ boson, the contributions are the same order, due to the additional $\epsilon^2$ suppression due to the $Z^\prime$ propagator.

There are no further contributions to be considered from the other neutral gauge bosons. Indeed,
for the photon $A$, we see from Eq.~\eqref{eq:A} that the interaction with down-type quarks is proportional to the identity matrix in flavour space, so that there are no FCNC from the photon interaction.
Concerning $W^{4,5}$, we see from Eqs.~\eqref{eq:W4} and~\eqref{eq:W5} that these gauge bosons always couple
a SM particle with an exotic one in the interaction basis. In order to obtain a $W^{4,5}$-mediated $b\to s$, we need to consider the interaction with one of the exotic interaction eigenstates, which will contain a SM mass eigenstate due to the rotation matrix $V^{(d)}$. As indicated earlier, this occurs only at order $O(\epsilon)$. Furthermore, the process is mediated by a heavy gauge boson, adding a further $O(\epsilon^2)$ suppression. Therefore the $W^{4,5}$ contributions to the process are of order $O(\epsilon^3)$ and  can be neglected compared to the $O(\epsilon^2)$ NP contributions from $Z$ and $Z'$ gauge bosons.

\subsection{Wilson coefficients and lepton-flavour violation}

 The joint effect of the two $O(\epsilon^2)$ contributions from $Z^\prime$ and $Z$ processes in our 331 model can be rewritten introducing the quantities \begin{equation}
\begin{split}
	f^{Z'}=&-\frac{1}{2\sqrt{2}G_F V_{tb}V_{ts}^*}\frac{4\pi}{\alpha}\frac{1}{3-\tan^2\theta_W}\frac{g^2}{M^2_{Z'}}V^{(d)*}_{3k}V^{(d)}_{3l}\\
	f^{Z}=&-\frac{1}{2\sqrt{2}G_F V_{tb}V_{ts}^*}\frac{4\pi}{\alpha}\frac{1 
	}{8}
	\frac{g^2}{M^2_{Z}}\sum_\lambda \hat{V}^{(d)*}_{\lambda k}\hat{V}^{(d)}_{\lambda l}\\
	\lambda^{(L)}_{ij}=&V^{(e)*}_{1i}V^{(e)}_{1j} \qquad\qquad \lambda^{(R)}_{ij}=W^{(e)*}_{3i}W^{(e)}_{3j}
\label{eq:factors}
\end{split}
\end{equation}
where $\theta_{331}$ and $g_X$ have been expressed in terms of $\theta_W$ and $g$ by using Eq.~\eqref{eq:wein}.  In order to focus on $b\to s$ transitions,
let us set the quark indices to $k=2$ and $l=3$
and rename coefficients and operators by removing the corresponding labels.  We get
\begin{equation}
	\mathcal H_{\text{eff}}\supset C^{ij}_9O^{ij}_9+C^{ij}_{10}O^{ij}_{10}
\end{equation}
where the operators $O^{ij}_{9,10}$ denote operators with given lepton flavours $i,j$, with the same normalisation as in Eq.~\eqref{eq:OP}.
We obtain the following NP contributions to the Wilson coefficients
\begin{equation}
\begin{split}
	C^{ij}_9&=f^{Z'}\Bigl[-\frac 1 2 \lambda^{(L)}_{ij}+\frac{1-2\tan^2\theta_W}2\lambda^{(R)}_{ij}+\\&+\frac{1+\tan^2\theta_W}4 \delta_{ij}\Bigr]+f^{Z}(-1+3\tan^2\theta_W)\delta_{ij}
\label{eq:system1}
\end{split}
\end{equation}
\begin{equation}
\begin{split}
	C^{ij}_{10}&=f^{Z'}\Bigl[\frac 1 2 \lambda^{(L)}_{ij}+\frac{1-2\tan^2\theta_W}2\lambda^{(R)}_{ij}+\\&+\frac{-1+3\tan^2\theta_W}4 \delta_{ij}\Bigr]+f^{Z}(1+\tan^2\theta_W)\delta_{ij}
\label{eq:system2}
\end{split}
\end{equation}
We see that LFUV contributions arise from the $Z'$ contribution, whereas the $Z$ contribution does not depend on the lepton flavour. In addition to the violation of lepton-flavour universality, our model allows for lepton-flavour violation, such as $b\to s \ell'^+\ell^-$  for $\ell'\neq\ell$. However, since there have been no experimental indications of such processes up to now, we will assume that these processes are suppressed, and for simplicity, we will set these coefficients to 0 when the two lepton indices are different, for any $i\neq j$. Imposing this, we get the system
\begin{equation}
\begin{cases}
	f^{Z'}\Bigl[- \lambda^{(L)}_{ij}+(1-2\tan^2\theta_W)\lambda^{(R)}_{ij}\Bigr]=0\\
	f^{Z'}\Bigl[ \lambda^{(L)}_{ij}+(1-2\tan^2\theta_W)\lambda^{(R)}_{ij}\Bigr]=0
\end{cases}\qquad \text{if } i\neq j;
\end{equation}
The trivial solution $f^{Z'}=0$ has to be discarded, since it would remove the only source of LFUV, i.e. the coupling of the charged leptons to $Z'$. The alternative solution is
\begin{equation}
	\lambda^{(L)}_{ij}=\lambda^{(R)}_{ij}=0, \qquad \text{if } i\neq j.
\label{lambda:st0}
\end{equation}
Due to the definitions of $\lambda^{(L,R)}_{ij}$ in Eq.~\eqref{eq:factors}, this solution implies that $V^{(e)}_{1I}$ can be nonzero for a single index $I$ among 1,2,3, and the same holds for a single $J$ among 1,2,3 for $W^{(e)}_{3J}$ \footnote{Assuming e.g. $I=1$, that is $V^{(e)}_{11}\neq 0$, Eq. \eqref{lambda:st0} implies $V^{(e)*}_{11}V^{(e)}_{12}= V^{(e)*}_{11}V^{(e)}_{13}=0$, that is $V^{(e)}_{12}=0$ and $V^{(e)}_{13}=0$.}. In other words, we require that the left-handed interaction eigenstate of the first generation and the right-handed interaction eigenstate of the third generation are also mass eigenstates.
Due to the unitarity of these $5\times 5$ matrices, we have then
\begin{eqnarray}\label{eq:lambdadef}
\lambda^{(L)}_{I}&\equiv& \lambda^{(L)}_{II}=|V^{(e)}_{1I}|^2=1-|V^{(e)}_{14}|^2-|V^{(e)}_{15}|^2 \\
\lambda^{(R)}_{J}&\equiv & \lambda^{(R)}_{JJ}=|W^{(e)}_{3J}|^2=1-|W^{(e)}_{34}|^2-|W^{(e)}_{35}|^2
\label{eq:lambdadef2}
\end{eqnarray}
which means that they must both stay within the [0,1] range, keeping in mind that $V$ and $W$ entries on the right hand side of Eqs.~(\ref{eq:lambdadef})-(\ref{eq:lambdadef2}) are of order $\epsilon$.  In the following, and for simplicity of notation, repeated indices (like $II$ or $ee$) will be denoted with a single index ($I$ or $e$).

We now consider two different scenarios:
\begin{itemize}
	\item Case A: the index $I$ for which the rotation matrix element $V^{(e)}_{1I}$ is nonzero is the same as the index $J$ for which the element $W^{(e)}_{3J}$ is non-vanishing;
	\item Case B: the two indices corresponding to non-vanishing matrix elements are different.
\end{itemize}

\subsubsection{Case A}

If we denote with $J$ the generation for which both entries for the rotation matrices are nonzero, we get
\begin{equation}
\begin{split}
	C^J_9&=f^{Z'}\Bigl[-\frac 1 2 \lambda^{(L)}_J+\frac{1-2\tan^2\theta_W}2\lambda^{(R)}_J+\frac{1+\tan^2\theta_W}{4}\Bigr]+\\&+f^{Z}(-1+3\tan^2\theta_W)\\
	C^J_{10}&=f^{Z'}\Bigl[\frac 1 2 \lambda^{(L)}_J+\frac{1-2\tan^2\theta_W}2\lambda^{(R)}_J+\frac{-1+3\tan^2\theta_W}4 \Bigr]+\\&+f^{Z}(1+\tan^2\theta_W)
\end{split}
\end{equation}
We get identical Wilson coefficients for the other two generations $i\neq J$, for which the entries in the rotation matrices vanish,
\begin{equation}
\begin{split}
	&C^i_9=f^{Z'}\frac{1+\tan^2\theta_W}4+f^{Z}(-1+3\tan^2\theta_W)\\
	&C^i_{10}=f^{Z'}\frac{-1+3\tan^2\theta_W}4+f^{Z}(1+\tan^2\theta_W).
\end{split}
\end{equation}
Inverting these relations we get
\begin{equation}
\begin{split}
	&f^{Z'}=\frac{(1+\tan^2\theta_W)C^i_9+(1-3\tan^2\theta_W)C^i_{10}}{2\tan^2\theta_W(1-\tan^2\theta_W)}\\
	&f^{Z}=\frac{(1-3\tan^2\theta_W)C^i_9+(1+\tan^2\theta_W)C^i_{10}}{8\tan^2\theta_W(1-\tan^2\theta_W)}\\
	&\lambda^{(L)}_Jf^{Z'}=C^i_9-C^i_{10}-C^J_9+C^J_{10}\\
	&\lambda^{(R)}_Jf^{Z'}=\frac{C^i_9+C^i_{10}-C^J_9-C^J_{10}}{-1+2\tan^2\theta_W}.
\label{eq:solut}
\end{split}
\end{equation}
We have now to identify whether the electron corresponds to the index $J$ or not. As discussed in Sect.~\ref{statementoftheproblem}, we set to zero the corresponding NP contributions to the effective Hamiltonian, $C^e_{9,10}$, on the basis of phenomenological constraints.
\begin{itemize}
\item If we identify the electron with another index $i\neq J$ (identifying the electron with a generation with vanishing entries), we must have $C^i_{9,10}=0$. From Eq.~\eqref{eq:solut}, we obtain that
$f^{Z'}=0$, so that no LFUV could be generated. We have thus to discard this possibility.
\item If we identify the electron with the index $J$ (identifying the electron with the generation with a non-vanishing entry), we set the corresponding NP Wilson coefficients to zero.
In this case, Eq.~\eqref{eq:solut} yields constraints on the possible values for the muon Wilson coefficients $C^i_{9,10}=C^\mu_{9,10}$ (also equal to  $C^\tau_{9,10}$):
\begin{equation}\begin{split}\label{eq:caseA}
&C^\mu_{10}=C^\mu_{9} \times \frac{2\tan^2\theta_W(\tan^2\theta_W-1)+\lambda^{(L)}_e(\tan^2\theta_W+1)}
  {2\tan^2\theta_W(\tan^2\theta_W-1)+\lambda^{(L)}_e(3\tan^2\theta_W-1)}\\
&C^\mu_{10}=-C^\mu_{9} \times \\&\frac{2 \tan^2 \theta_W(\tan^2\theta_W-1)+\lambda^{(R)}_e(2\tan^4\theta_W+\tan^2\theta_W-1)}
  {2  \tan^2 \theta_W(\tan^2\theta_W-1)-\lambda^{(R)}_e(6\tan^4\theta_W-5\tan^2\theta_W+1)}
\end{split}\end{equation}
Since $0\leq \lambda^{(L)}_e, \lambda^{(R)}_e\leq 1$, these expressions yield a wedge in the $(C^\mu_{9},C^\mu_{10})$ plane. The constraint from $\lambda^{(L)}$ is the more stringent one, imposing the ratio $C^\mu_{10}/C^\mu_{9}$ to remain between -1.75 and -1 (we use $\sin^2\theta_W\simeq 0.235$), as indicated as a grey wedge on the top part of  Fig.~\ref{fig:plane}.
 \end{itemize}

In summary, in case A, we find that the electron has to be identified with the generation with a non-vanishing entry in the rotation matrices $V$ and $W$. Muons and taus give the same NP contribution to the Wilson coefficients $C_9$ and $C_{10}$
Eqs.~\ref{eq:caseA}, imposing that $|C^\mu_{10}|\geq |C^\mu_{9}|$.

\begin{figure}
\centering
\includegraphics[width=.7\columnwidth]{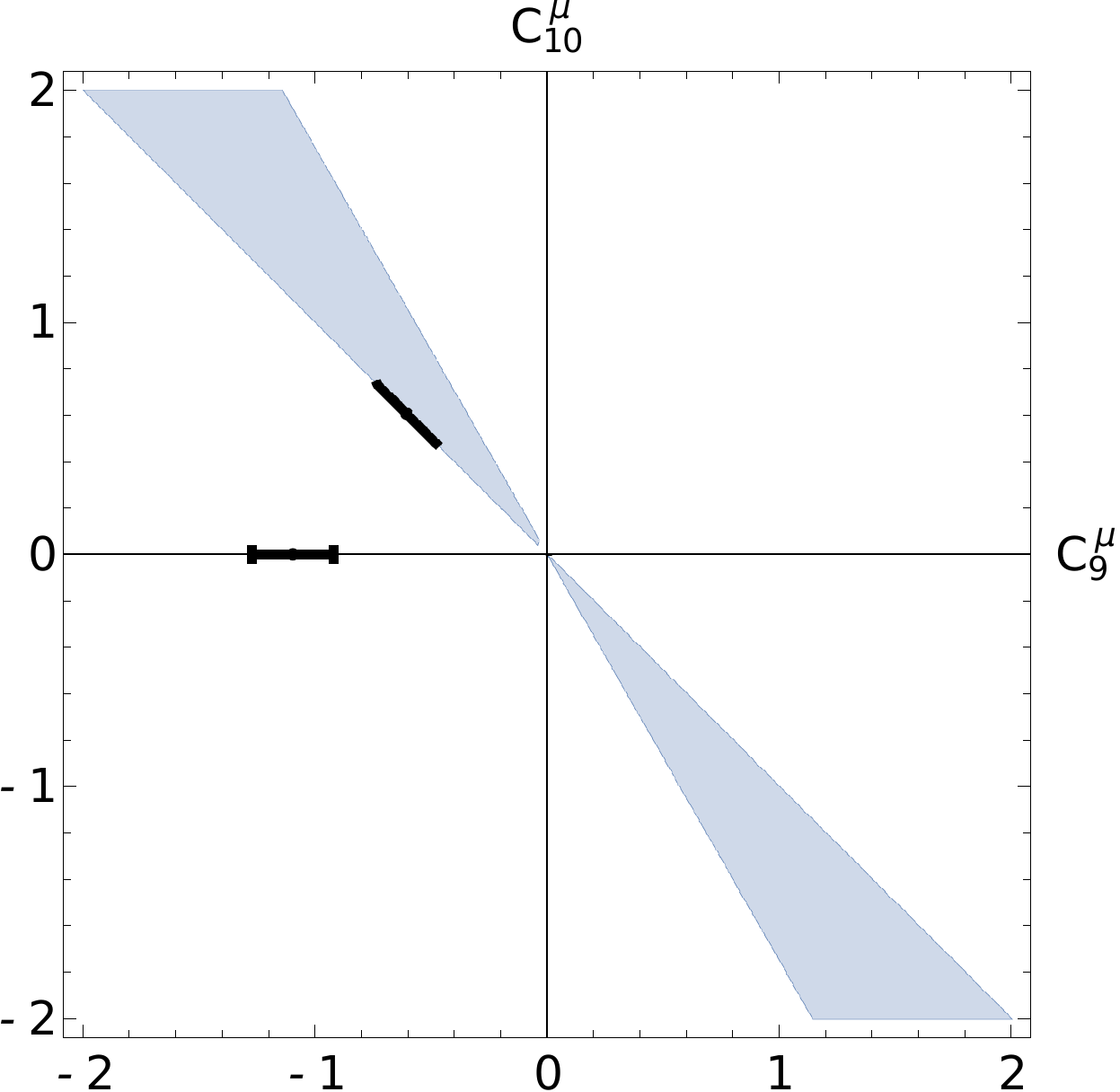}
\includegraphics[width=.7\columnwidth]{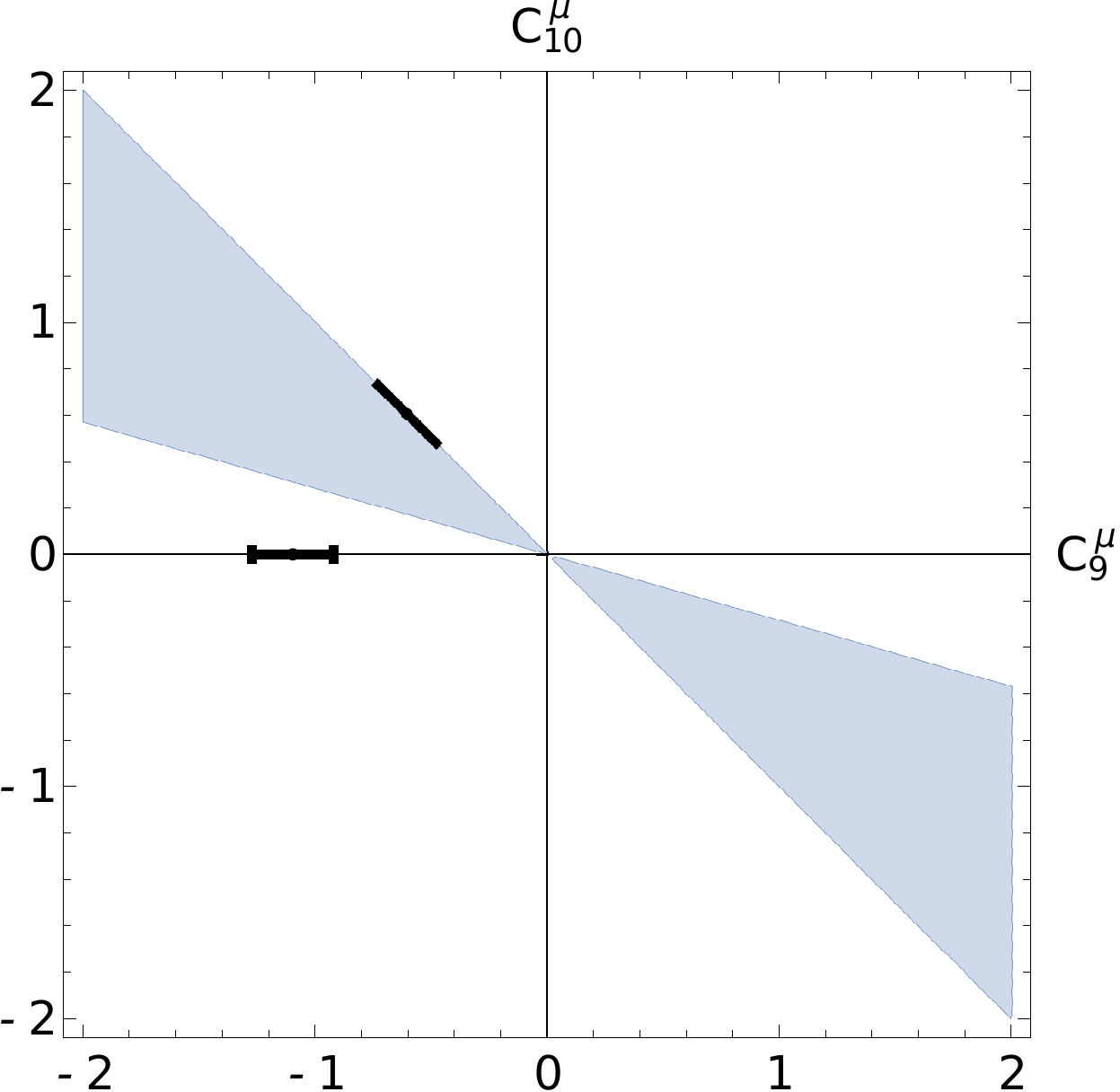}
\caption{Regions allowed for the Wilson coefficient $C^\mu_{9}$ and $C^\mu_{10}$ (abscissa and ordinate, respectively) in scenarios A (top) and B (bottom). The thick black intervals correspond to the 1$\sigma$ interval for one-dimensional scenarios from Ref.~\cite{Capdevila:2017bsm}.}
\label{fig:plane}
\end{figure}

\subsubsection{Case B}

In case B, we have two different indices $I\neq J$ such that $V^{(e)}_{1I}\neq0$ and $W^{(e)}_{3J}\neq0$ (so that $\lambda^{(L)}_I\neq 0$ and $\lambda^{(R)}_J\neq 0$). The system of equations defining the Wilson coefficients, Eqs. \eqref{eq:system1}\eqref{eq:system2}, becomes
\begin{equation}
\begin{cases}
	&C^I_9=f^{Z'}\Bigl[-\frac 1 2 \lambda^{(L)}_I+\frac{1+\tan^2\theta_W}4\Bigr]+f^Z(-1+3\tan^2\theta_W)\\
	&C^I_{10}=f^{Z'}\Bigl[\frac 1 2 \lambda^{(L)}_I+\frac{-1+3\tan^2\theta_W}4 \Bigr]+f^Z(1+\tan^2\theta_W)\\
	&\begin{aligned}[l]C^J_9=&f^{Z'}\Bigl[\frac{1-2\tan^2\theta_W}{2}\lambda^{(R)}_J+\frac{1+\tan^2\theta_W}4 \Bigr]+\\&+f^Z(-1+3\tan^2\theta_W)\end{aligned}\\
	&\begin{aligned}[l]C^J_{10}=&f^{Z'}\Bigl[\frac{1-2\tan^2\theta_W}{2}\lambda^{(R)}_J+\frac{-1+3\tan^2\theta_W}4 \Bigr]+\\&+f^Z(1+\tan^2\theta_W)\end{aligned}
\end{cases}
\end{equation}
Inverting with respect to $f^{Z'}, f^Z, \lambda^{(L)}_Jf^{Z'}, \lambda^{(R)}_Jf^{Z'}$ we get
\begin{equation}
\begin{split}
	f^{Z'}&=\frac{C^I_9+C^I_{10}}{2\tan^2\theta_W}+\frac{C^J_9-C^J_{10}}{1-\tan^2\theta_W}\\
	f^Z&=\frac{C^I_9+C^I_{10}}{8\tan^2\theta_W}+\frac{-C^J_9+C^J_{10}}{4(1-\tan^2\theta_W)}\\
	\lambda^{(L)}_If^{Z'}&=-C^I_9+C^I_{10}+C^J_9-C^J_{10}\\
	\lambda^{(R)}_Jf^{Z'}&=\frac{-C^I_9-C^I_{10}+C^J_9+C^J_{10}}{1-2\tan^2\theta_W}
\label{eq:solutions}
\end{split}
\end{equation}
Moreover, if we denote $K$ the remaining SM generation ($K\neq I,J$) we have the following relationships
\begin{equation}\begin{split} \label{eq:WC3gen}
C^K_9 &=\frac{1}{2}[C^I_9+C^I_{10}+C^J_9-C^J_{10}]\\
C^K_{10} &=\frac{1}{2}[C^I_9+C^I_{10}-C^J_9+C^J_{10}]\end{split}
\end{equation}

We still have not identified which of the $I,J,K$ indices refers to the electron, muon, or tau leptons:
\begin{itemize}
\item If we identify the electron with $J$, we set $C^J_{9}=C^J_{10}=0$ and from the first and last relations of Eq.~\eqref{eq:solutions} we get
\begin{equation}
	\lambda^{(R)}_J=-\frac{2\tan^2\theta_W}{1-2\tan^2\theta_W}<0
\end{equation}
leading to an inconsistency, since the $\lambda$ must be non-negative.
\item If we identify the electron with $K$, we set $C^K_{9}=C^K_{10}=0$ and from Eq.~\eqref{eq:WC3gen} we get
\begin{equation}\begin{split}
C_{9}^I&=-C_{10}^I=-\frac{1}{2}f^{Z'}\lambda^{(L)}_I \\ C_{9}^J&=C_{10}^J=\frac{1}{2} f^{Z'}(1-2\tan^2\theta_W)\lambda_J^{(R)}
\end{split}\end{equation}
which can be used in Eq.~(\ref{eq:solutions}) to show that $f^{Z}=f^{Z'}=0$, so that this solution can be discarded.
\item If we identify the electron with $I$, we set $C^I_{9}=C^I_{10}=0$, the solutions Eq.~\eqref{eq:solutions} become
\begin{equation}
\begin{split}
	f^{Z'}&=\frac{C^J_9-C^J_{10}}{1-\tan^2\theta_W}\\
	f^Z&=\frac{-C^J_9+C^J_{10}}{4(1-\tan^2\theta_W)}\\
	\lambda^{(L)}_If^{Z'}&=C^J_9-C^J_{10}\\
	\lambda^{(R)}_Jf^{Z'}&=\frac{C^J_9+C^J_{10}}{1-2\tan^2\theta_W}
\label{startfromthis}
\end{split}
\end{equation}
from which we can read the expressions for the $\lambda$:
\begin{equation}\begin{split}
\lambda^{(L)}_I&=1-\tan^2\theta_W\in[0,1]\\
	\lambda^{(R)}_J&=\frac{C^J_9+C^J_{10}}{C^J_9-C^J_{10}}\,\frac{1-\tan^2\theta_W}{1-2\tan^2\theta_W}.
\end{split}\end{equation}
leading to the following conditions on the non-vanishing NP Wilson coefficients
\begin{equation}\begin{split}\label{eq:caseB}
C^J_{10}&=-C^J_9\times\frac{1-\tan^2\theta_W+(2\tan^2\theta_W-1)\lambda^{(R)}_J}{1-\tan^2\theta_W-(2\tan^2\theta_W-1)\lambda^{(R)}_J}\\
C^K_{9}&=-C^K_{10}=\\&=C^J_9\times \frac{\tan\theta_W^2-1}{\tan\theta_W^2-1+(2\tan^2\theta_W-1)\lambda^{(R)}_J}\end{split}
\end{equation}
We see that the value found for $\lambda^{(L)}_I=\lambda^{(L)}_e$ lies in the allowed interval $[0,1]$. Furthermore, requiring that $\lambda^{(R)}_J$ also remains in this interval yields a constraint on the Wilson coefficients: if we identify the muon with $K$, we have the exact equality $C^\mu_{10}/C^\mu_9=-1$, and if we identify the muon with $J$, the slope $C^\mu_{10}/C^\mu_9$ is constrained between -1 and -0.28 (using $\sin^2\theta_W\simeq 0.235$). These constraints are indicated in grey on the bottom part of  Fig.~\ref{fig:plane}.
\end{itemize}

In summary, in case B, we find that the electron generation must be identified with the non-vanishing entry $I$  in the rotation matrices $V$. Two possibilities can be considered concerning the non-vanishing entry $J$ in the rotation matrices $W$.
If we identify $J$ with the muon generation,
muons and taus have different NP contribution for the corresponding Wilson coefficients $C_9$ and $C_{10}$, imposing that $|C^\mu_{10}|\leq |C^\mu_{9}|$, the NP contribution to $C^\mu_{10}$ is different from zero, and $C_9^\tau=-C_{10}^\tau$.
If we identify $J$ with the tau generation, one gets again  different NP contributions for the Wilson coefficients $C_9$ and $C_{10}$ for muons and taus,  the roles played by muons and taus are reversed, and thus one gets $C_9^\mu=-C_{10}^\mu$.  Both cases yield thus NP contributions given by Eq.~\eqref{eq:caseB}.

\section{Comparison with global analyses}
\label{sec:global}

We perform a comparison between the 331 model contributions to the process $b\to s \ell^+ \ell^-$ and the global analysis of $b\to s\ell \ell$ anomalies performed in Refs.~\cite{sdg, sdg2,Capdevila:2017bsm} (similar results were obtained in recent works from other groups, see Refs.~\cite{Capdevila:2017bsm, Altmannshofer:2017yso, Geng:2017svp, Ciuchini:2017mik, Hurth:2017hxg}). In these works, the authors pointed out scenarios in which NP contributions to the Wilson coefficients $C_{9('),10}^\mu$ are favoured whereas no NP contributions occur for other Wilson coefficients (including all the electronic ones). In particular they identified three specific one-dimensional scenarios as particularly favoured:
\begin{itemize}
	\item NP in $C_9^\mu=-C_{9'}^\mu$, with the 1$\sigma$ interval [-1.18, -0.84]: this scenario cannot be described in the framework of our non-minimal 331 model, where no FCNC arise for right-handed quarks, meaning that $C_{9'}^\mu=0$ (see Sec.~\ref{sec:gaugebosoncontributions});
	\item NP in $C_9^\mu$, within the 1$\sigma$ interval [-1.27, -0.92]. From the discussion of the previous section and Fig.~\ref{fig:plane}, we observe that this scenario is allowed neither in scenario A nor B.
	\item NP in $C_9^\mu=-C_{10}^\mu$, within the 1$\sigma$ interval [-0.73, -0.48]. From the discussion of the previous section and Fig.~\ref{fig:plane}, we see that this scenario is allowed in both scenarios A and B.
\end{itemize}
Our non-minimal 331 model appears to be able to account for the $b\to s\ell\ell$ anomalies observed as far as we consider the $C_9^\mu=-C_{10}^\mu$ case. More generally, it would be able to reproduce other favoured values for the two-dimensional scenario $(C_9^\mu,C_{10}^\mu)$ with negative NP contributions to $C_9^\mu$ and positive to $C_{10}^\mu$ (see top-left plot in Fig.~1 in Ref.~\cite{Capdevila:2017bsm}).

For simplicity and illustration of the potential of our 331 model, we will focus here on the one-dimensional (1D) scenario $C_9^\mu=-C_{10}^\mu$ considered in Refs. \cite{sdg, Capdevila:2017bsm}. Imposing this equality, we see that in both cases A and B we have $\lambda_e^{(L)}=1-\tan^2\theta_W$~\footnote{According to Eq.~(\ref{eq:lambdadef}), $\lambda_I^{(L)}-1=O(\epsilon^2)$, indicating that $\epsilon$ should be of the same order of magnitude as $\tan\theta_W$ in this scenario. Nevertheless, this estimate can be relaxed by the magnitude of the lepton Yukawa couplings, on which $\lambda_I^{(L)}$ depends.} and
\begin{equation}\begin{split}\label{eq:wilson-toconstrain}
C_9^\mu&=-C_{10}^\mu=f^{Z'}\frac{1-\tan^2\theta_W}{2}=\\
  &=-\frac{1}{V_{tb}V_{ts}^*}\frac{1-\tan^2\theta_W}{3-\tan^2\theta_W}\frac{4\pi}{\alpha}\frac{M_W^2}{M_{Z'}^2}V_{3k}^{(d)*}V_{3l}^{(d)}\qquad [1D]\end{split}
\end{equation}
so that NP contribution to $C_9^\mu$ is given by parameters of the 331 model included in $f^{Z'}$, where the only unknown quantities are $M_{Z'}$ and $V^{*(d)}_{32}V^{(d)}_{33}$. These can be further constrained by other processes, and in particular $B_s$ meson mixing, as  explained in the next section.

{\begin{figure}\centering
	\subfloat[][~]{
	\scalebox{0.7}{\includegraphics{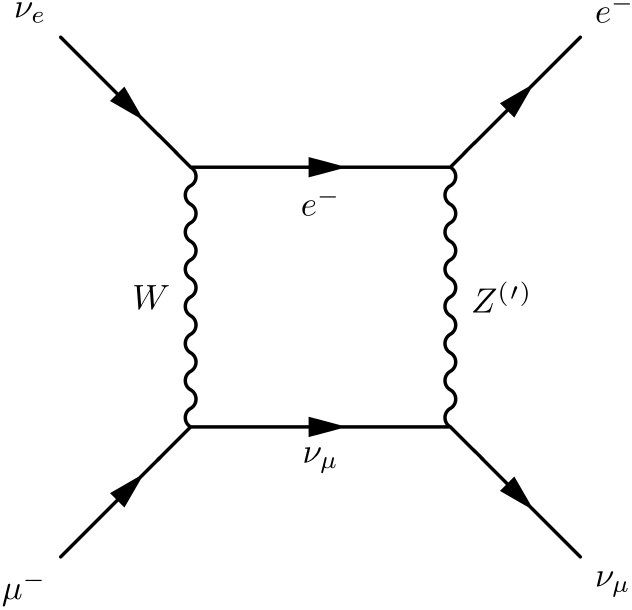}}
	}
	\vspace{0.5cm}
	\subfloat[][~]{
	\scalebox{0.7}{\includegraphics{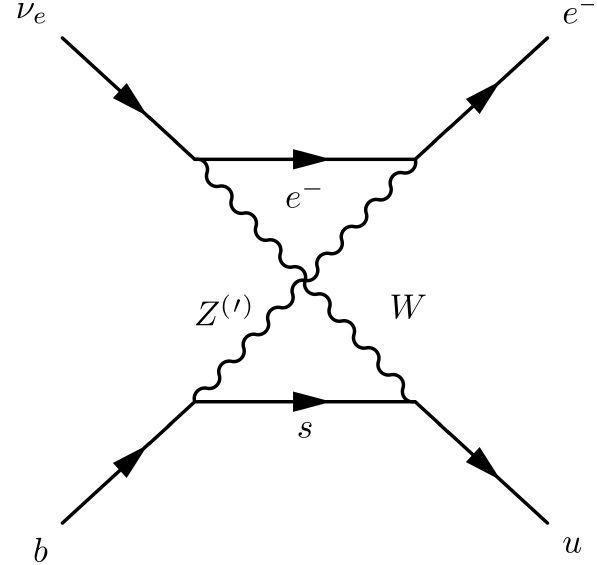}}
	}\qquad\qquad
	\vspace{0.5cm}
	\subfloat[][~]{
	\scalebox{0.7}{\includegraphics{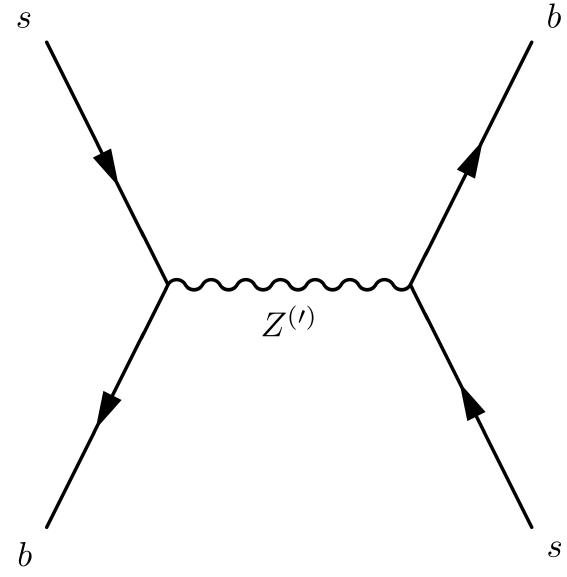}}
	}
\caption{Gauge contributions to the violation of unitarity of the CKM matrix in the first row (for matrix elements determined leptonic and semileptonic processes) and to $B_s\bar{B}_s$ mixing (see Refs.~\cite{Gauld:2013qba,Gauld:2013qja}). }
\label{fig:Bsmix}
\end{figure}}

\section{Phenomenological constraints on $Z$ and $Z'$ couplings}\label{sec:mixing}

We have built our 331 model in order to generate vector/axial LFUV contributions to $b\to  s\ell\ell$ transitions. This has led us to assume that the dominant contributions for these couplings ($bs$ and $\mu\mu$) came from the gauge bosons rather than the Higgs sector, and actually that the dominant contributions came from anomalous couplings of the $Z$ gauge boson as well as tree-level exchanges of a $Z'$ gauge boson. Even in this restricted setting, there are additional constraints to be considered on these couplings from the phenomenological point of view, as discussed in Refs.~\cite{Gauld:2013qba,Gauld:2013qja,331:buras,331:burasmu,331:burasZZ}.

A first class of constraints for additional contributions from neutral gauge bosons comes from the violation of unitarity in the CKM matrix. One has to consider the corrections to the decay $\mu^-\to e^-\nu_\mu\bar\nu_e$ (as it defines the normalisation for all decays through $G_F$) as well as the decays $b,s,d\to u e^-\bar\nu_e$ (leading to $|V_{ub}|$, $|V_{us}|$ and $|V_{ud}|$ determinations assuming the SM). This corresponds to box diagrams involving both $W$ and $Z$ or $Z'$ bosons, as shown in graphs (a) and (b) of Fig.~\ref{fig:Bsmix}. One can expect the $Z'$ contribution to be small, as the diagrams require to have a $Z'$ coupling to the first generation, which is suppressed in our model. On the other hand, the FCNC couplings of the $Z$ to quarks occur (in principle) between all down-type quarks, meaning that we need a detailed understanding of the $O(\epsilon)$ mixing matrix $\hat{V}^{(d)}$, see Eq.~\eqref{eq:Zcontr1}, in order to compute this  correction in our model. Such a detailed knowledge might be obtained by a complete analysis of all flavour constraints on our model, which is far beyond the scope of the present article.

A second constraint comes from $B_s-\overline{B}_s$ mixing to which both
$Z$ and $Z'$ gauge bosons give a tree level contribution, as can be seen on Fig.~\ref{fig:Bsmix}. This constraint can thus provide useful information in addition to the $b\to s\ell\ell$ decay.
As before, we restrict our discussion to contributions of order ${\mathcal O}(\epsilon^2)$, borrowing from the results in  Sec.~\ref{sec:gaugebosoncontributions}. At this order, $Z$ gives no contributions to the mixing. Indeed, the $bsZ$ vertex has a suppression of ${\mathcal O}(\epsilon^2)$, due to the structure of the unitary matrices needed to obtain physical states. The contribution to $B_s-\overline{B}_s$ mixing will have two such vertices, and hence be suppressed by a factor ${\mathcal O}(\epsilon^4)$.
Concerning the $Z'$ contribution, we only need to take into account the ${\mathcal O}(\epsilon^2)$ suppression coming from the heavy gauge boson propagator, since the $bs$ vertex for this gauge boson is already mediated at $\mathcal O (\epsilon^0)$.

As discussed in App.~\ref{sec:ZZprimecouplings}, the relevant part of the interaction for $B_s-\overline{B}_s$ is thus (in the interaction eigenbasis)
{\begin{equation}
\centering
	\mathcal L_{Z'}\supset\frac {\cos\theta_{331}}{g_X}Z'_\mu\frac{g_X^2}{3\sqrt6\cos^2\theta_{331}}\bar D^L\gamma^\mu\begin{pmatrix}0&0&0\\0&0&0\\0&0&1\\\end{pmatrix} D^L
\label{BsBsbar}
\end{equation}
}
Expressing in terms of effective operators of eigenstates and using Eq.~\eqref{eq:wein}, one obtains
\begin{equation}
	\begin{split}
		&\mathcal H_{\rm eff}\supset \frac{g_X^2}{54M^2_{Z'}\cos^2 \theta_{331}}(V^{*(d)}_{3k}V^{(d)}_{3l})^2(\overline{D_k}\gamma^\mu D_l)(\overline{D_k}\gamma^\mu D_l)=\\
					&=\frac{8G_F}{\sqrt 2 (3-\tan^2\theta_W)}\frac{M_W^2}{M_{Z'}^2}(V^{*(d)}_{3k}V^{(d)}_{3l})^2(\overline{D_k}\gamma^\mu D_l)(\overline{D_k}\gamma^\mu D_l)\\
	\end{split}
\end{equation}
where we will focus as usual on the case $k=2,l=3$.

The SM contribution to the mixing reads~\cite{Lenz:2010gu}
\begin{equation}
	\mathcal H^\text{SM}_{\rm eff} = (V_{ts}^*V_{tb})^2\frac{G_F^2}{4\pi^2} M_W^2 \hat{\eta}_B S\Bigl(\frac{\overline{m_t}^2}{M_W^2}\Bigr)(\overline{s_L}\gamma^\mu b_L)(\overline{s_L}\gamma^\mu b_L)
\end{equation}
where $S$ is the Inami-Lim function and $\overline{m_t}$ is  the top quark mass defined in the
$\overline{MS}$ scheme. As in Ref.~\cite{Lenz:2010gu}, we take  $S\Bigl(\frac{\overline{m_t}^2}{M_W^2}\Bigr)\simeq 2.35$, for a top mass of about 165 GeV, and $\hat{\eta}_B=0.8393\pm 0.0034$, which comprises QCD corrections.

Considering the modulus of the ratio of the NP contribution over the SM, one gets
\begin{equation}\begin{split}
	r_{B_s}&=\left|\frac{C_\text{NP}}{C_\text{SM}}\right| =\\&= \frac{32\pi^2|V^{*(d)}_{32}V^{(d)}_{33}|^2}{\sqrt 2 (3-\tan^2\theta_W)|V_{ts}^*V_{tb}|^2G_FM_W^2\hat{\eta}_B S} \frac{M_W^2}{M_{Z'}^2}\end{split}
\end{equation}
In this expression, the only values that are not assigned are $d=V^{*(d)}_{32}V^{(d)}_{33}$ and $M_{Z'}^2$ or, equivalently, $\frac{M_W^2}{M_{Z'}^2}$. Since $d$ consists of products of elements of unitary matrices, its value must necessarily lie in the interval $[-1,1]$ (assuming that it is real).

In order to get an impression of the values allowed, we perform a scan varying $d$ in $[-1,1]$ and $M_W/M_{Z'}$ in the range $[0,0.1]$, corresponding roughly to a NP scale at least of the order of 10 times the electroweak scale. We allow the NP contributions to the $B_s$ mixing to be at most $10\%$ (i.e., $r_{B_s}\leq 0.1$), in agreement with recent global fits to NP in $B_d$ and $B_s$ mixings where the constraint from $\Delta M_s$ is the main limiting factor~\cite{Lenz:2012az,Charles:2013aka}.
For those values, we evaluate the NP contribution to the Wilson coefficient $C_9^\mu=-C_{10}^\mu$ in the one-dimensional scenario as expressed in Eq.~\eqref{eq:wilson-toconstrain}. The allowed values found in the scan are plotted in Fig.~\ref{fig:parabolic}.

We see that values of $C_9^\mu=-C_{10}^\mu$ can reach -0.6, in agreement with the results of global analyses of $b\to s\ell\ell$, corresponding to $r_{B_s}=0.1$, $M_W/M_{Z'}=0.1$ and $d\simeq -0.005$. The allowed region is limited by the fact that we have numerically
\begin{equation}\begin{split}
r_{B_s}&\simeq 347\cdot 10^3 \times  \left(\frac{M_W}{M_{Z'}}\right)^2 \times d^2\leq 0.1\\
C_9^\mu &\simeq 11.3 \cdot 10^3 \times \left(\frac{M_W}{M_{Z'}}\right)^2 \times d\qquad |d|\leq 1\end{split}
\end{equation}
using Refs.~\cite{Koppenburg:2017mad, Patrignani:2016xqp}, which
leads to the parabolic constraint $r_{B_s}=(C_9^{\mu})^ 2 \times 0.003/(M_W/M_{Z^\prime})^2\geq 0.3\times (C_9^{\mu})^ 2$, represented in Fig.~\ref{fig:parabolic}.

As we saw in the previous sections, our 331 model can accommodate various NP contributions to $(C_9^\mu,C_{10}^\mu)$. In the simple one-dimensional scenario $C_9^\mu=-C_{10}^\mu$, we can accommodate both $B_s\bar{B}_s$ mixing and $b\to s\ell\ell$ data, with a NP scale (and in particular a $Z'$) around the TeV scale. Choosing different values for $(C_9^\mu,C_{10}^\mu)$ would extend the parameter space for NP allowed, with the possibility to use not only  the value of $f^{Z'}$, but also $f^{Z}$, to accommodate the data.

\begin{figure}
\centering
\includegraphics[width=.95\columnwidth]{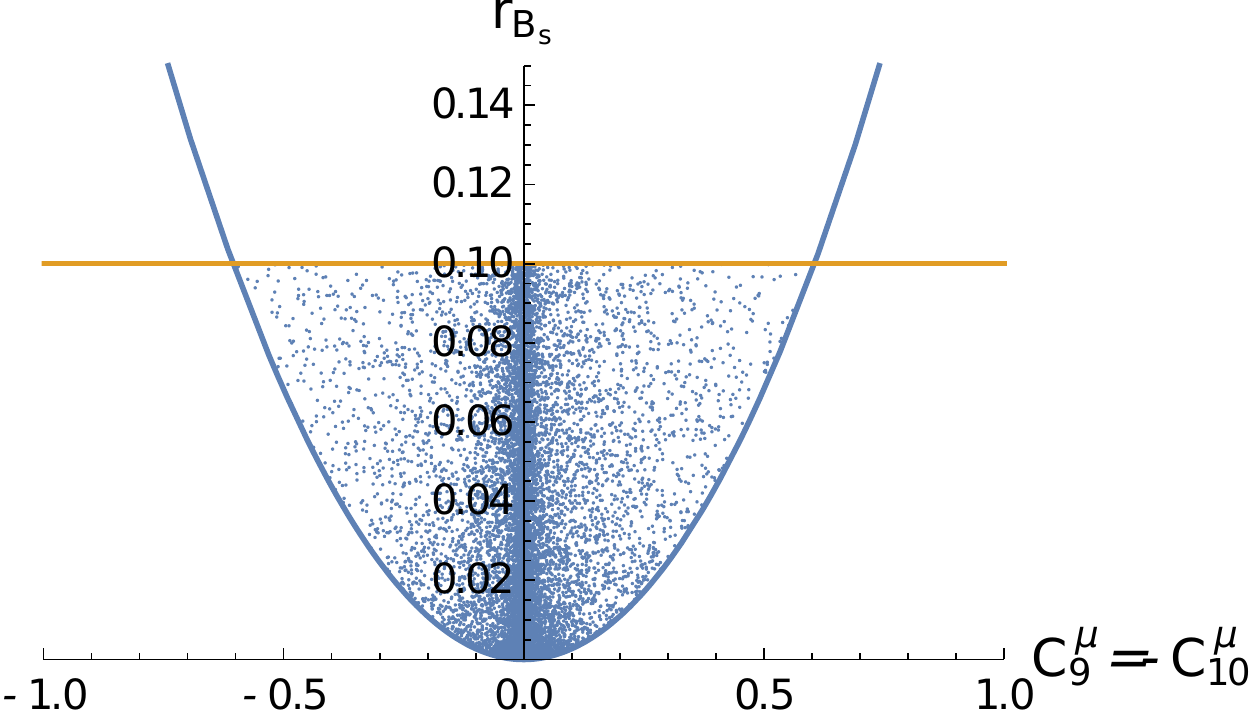}
\caption{Allowed points in the ($C^\mu_{9},r_{B_s}$) plane.}
\label{fig:parabolic}
\end{figure}

 A third kind of constraints comes from the study of contact interactions from the LEP data on $e^+e^-\to q\bar{q}$ or $\ell^+\ell^-$, as analysed in Ref.~\cite{Schael:2013ita} (Tables 3.14 and 3.15) and the LHC data on $pp$ collisions, for instance the ATLAS data~\cite{ATLAS:2017wce} and reanalysed in Table 1 of Ref.~\cite{Greljo:2017vvb}. These studies impose constraints on the couplings $\Delta$
introduced in Eq.~(\ref{eq:Delta}) as NP $O(\epsilon^2)$ operators of the effective Hamiltonian involving only light charged fermions and being mediated by charged currents. This means that the $\Delta$ couplings are generally of $O(0.01)$ or less.
A few general statements can be made even before studying these constraints in detail. Ref.~\cite{Schael:2013ita} uses $Z$-decays in order to put constraints on various kinds of patterns for the contact interactions, leading to NP ranging from 2 to 15 TeV, corresponding to upper bounds on the couplings $\Delta$ ranging from 0.15 to 0.003. The tables in Ref.~\cite{Greljo:2017vvb} lead to bounds on the couplings $\Delta$ ranging from 0.01 (in most of the case) down to  0.001 (for couplings concerning $u$- or $d$-quarks together with muons. This means that the structure of the matrix $V^{(u,d,e)}$ and $W^{(e)}$ must be moderately fine tuned (at the 10\% level) in order to accomodate both LEP and LHC bounds in general. A more detailed study of these constraints would require a thorough analysis of the patterns of deviations for all four-fermion operators, which is beyond the scope of this paper.

\section{Conclusions}\label{sec:outlook}

Among many achievements, the LHC experiments have been able to investigate many rare flavour processes, with very interesting outcomes.
In particular, the LHCb experiment has identified several deviations from the Standard Model in the $b\to s\ell\ell$ transitions, with interesting hints from violation of the lepton flavour universality. These deviations can be elegantly explained within model-independent effective approaches, where a few Wilson coefficients receive significant NP contributions. This has triggered a lot of theoretical work to identify viable models explaining such deviations, among which $Z'$ models and leptoquark models have been often used.

In the present paper, we try to embed a $Z'$ model in a more global extension, widely used in the literature, namely the 331 models where the gauge group $SU(3)_\text{C}\times SU(3)_\text{L}\times U(1)_X$ breaks down at a high scale into the SM gauge group, before undergoing a second transition at the electroweak scale. The minimal versions of such models do not feature lepton flavour universality violation as they have to obey anomaly cancellations. We thus investigated a non-minimal 331 model with 5 lepton triplets able to include LFUV. We described the choices made to build this model in order to have all additional gauge bosons and fermions with heavy masses (of the order of the scale of $SU(3)_\text{L}$ breaking) and electric charges similar to those present in the SM. We worked out how this model could reproduce the deviations observed in $b\to s\ell\ell$ transitions.
This requires us to assume that the deviations are dominated by neutral gauge boson contributions (anomalous $bsZ$ coupling
due to fermion mixing as well as flavour-changing neutral coupling to a heavy $Z'$ boson). The absence of significant contribution to $b\to see$ and lepton-universality violating processes allowed us to set constraints on the mixing matrices between interaction and mass fermion eigenstates.

We identified two different cases for the mixing matrices, with a rather simple outcome. Our model turns out to have no right-handed currents, but it is able to accommodate significant NP contributions to $C_9^\mu$ (negative) and $C_{10}^\mu$ (positive), in agreement with NP scenarios favoured by global fits. In each case, we could make predictions concerning the $\tau$ Wilson coefficients (the electron ones being assumed to receive no NP contribution). We considered additional phenomenological constraints on $Z$ and $Z'$ couplings in order to check the viability of our model: if the unitarity of the first row of the CKM matrix is not powerful in our case due to the large number of parameters involved, $B_s\bar{B}_s$ mixing proves much more powerful. 

Considering these results, it would be interesting to progress further in the study of this non-minimal 331 model. Since we are able to predict in each case the values of Wilson coefficients for $b\to s\tau\tau$ from the electronic and muonic ones, it would be interesting to predict the deviations arising to related observables from our model, whether in decays or in $B_s\bar{B}_s$ mixing~\cite{Grossman:1996qj,Bobeth:2011st,Alonso:2015sja,Kamenik:2017ghi}.

Under our simplifying assumptions (no lepton-flavour violation $b\to s\ell_i\ell_j$, no contribution to $b\to s ee$, opposite contribution to $C_9^\mu$ and $C_{10}^\mu$), we saw that we are able to accommodate both $b\to s\ell\ell$ and $B_s\bar{B}_s$ mixing observables at the price of a NP scale of order 1 TeV. Considering different values of NP contributions to $C_9^\mu$ and $C_{10}^\mu$ might also enable to increase the scale of NP allowed. It would also be interesting to compare this constraint with direct searches for $Z'$ bosons, taking into account the pattern of couplings specific to our model.  A first look at the constraints on contact interactions suggest that these bounds could be accommodate through a moderate fine tuning of the unitary matrices connecting mass and interaction eigenstates, but a more thorough analysis would naturally be very useful.

Moreover, it would also be natural to consider the other hints of LFUV currently present in flavour physics, namely $R_{D}$ and $R_{D^*}$. Global model-independent analyses show that the LFUV deviations seen in $b\to c\ell\nu$ branching ratios can be explained by vector/axial exchanges, whereas scalar/pseudoscalar exchanges are disfavoured~\cite{Freytsis:2015qca,Watanabe:2017mip}. In our model, the situation is a bit different compared to $b\to s\ell\ell$ transitions. Indeed the heavy charged bosons have no couplings with SM fields in the interaction eigenbasis, which means that the SM quark and lepton couplings will be induced again by mixing (each counting at ${\mathcal O}(\epsilon)$) and further suppressed by the heavy gauge boson mass, leading to a contribution ${\mathcal O}(\epsilon^4)$. The light $W^\pm$ bosons have diagonal couplings in the SM subspace in the interaction eigenbasis (see Eq.~\eqref{eq:W}), which means that LFUV will appear only due to mixing effects in leptons.

This effect can in principle be of order ${\mathcal O}(\epsilon^2)$ or lower, depending on the structure of the mixing in the neutral lepton sector. For this reason, the deviations observed in $b\to c$ transitions could also be explained in our model through gauge boson contributions only. The discussion requires an accurate analysis of the neutrino spectrum, and we leave it for future work.

The additional requirements from $R_{D}$ and $R_{D^*}$ would thus allow us to further refine our non-minimal 331 model, and to determine if it constitutes a viable alternative to explain the LFUV processes currently observed in $b$-decays. If it passes these tests, it could provide an interesting alternative to current NP models used to explain the deviations in $b$-quark decays, with a potential to be tested both through deviations in flavour processes among other generations of quarks and leptons and through
direct production searches at LHC.
\begin{acknowledgments}
We thank Monika Blanke for valuable comments on the manuscript.
SDG thanks  Universit\`{a} di Napoli Federico II and I.N.F.N. Sezione di Napoli for their kind hospitality during his stay where part of the present work was carried out. MM is thankful to the Erasmus Traineeship Program of  Universit\`{a} di Napoli Federico II and Universit\'{e} Paris Sud, and to the Laboratoire de Physique Th\'{e}orique d'Orsay for hosting her internship during which part of the work was carried out.
 This work received financial support from the grant FPA2014-61478-EXP and from the EU Horizon 2020 program from the grants No 690575, No 674896 and No. 692194 [SDG], from the DFG-funded Doctoral School ``KSETA'' [MM], from MIUR under Project No. 2015P5SBHT and from the INFN research initiative ENP [GR].
\end{acknowledgments}
\appendix
\section{Fermionic content of the model}\label{app:fermioniccontent}

We summarise the $U(1)$ charges of the fermionic content of our model (for the charged fermions) in Tab.~\ref{tab:fermi}. We recall
that the lower case letters denote light fields corresponding to the SM, whereas upper case letters correspond to heavy exotic fields. As discussed in Sect. \ref{choicebeta}, all fields have only charges already present in the SM.

\begin{table}[t]
\[
\begin{array}{ccrr}
\toprule
&\text{Fermion} & Q&X \\
\midrule
\multirow{12}*{Quarks}
&u^L_1, u^L_2	&	\frac 2 3 	&	0	\\
&d^L_1, d^L_2	&	-\frac 1 3 	&	0	\\
&u^R_1, u^R_2	&	\frac 2 3 	&	\frac 2 3	\\
&d^R_1, d^R_2	&	-\frac 1 3 	&	-\frac 1 3 	\\
&u^L_3	&	\frac 2 3 	&	\frac 1 3	\\
&d^L_3	&	-\frac 1 3 	&	\frac 1 3 	\\
&u^R_3	&	\frac 2 3 	&	\frac 2 3	\\
&d^R_3	&	-\frac 1 3 	&	-\frac 1 3 	\\
&B^L_{1,2}	&	-\frac 1 3 	&	0 	\\
&B^R_{1,2}	&	-\frac 1 3 	&	-\frac 1 3	\\
&T^L_3	&	\frac 2 3 	&	 \frac 1 3 	\\
&T^R_3	&	\frac 2 3 	&	 \frac 2 3 	\\
\midrule
\multirow{16}*{Leptons}
&e^{-L}_1	&	-1 	&	 -\frac 2 3 	\\
&e^{-R}_1	&	-1 	&	 -1	\\
&\nu^{L}_1	&	0 	&	 -\frac 2 3 	\\
&E^{-L}_1	&	-1 	&	 -\frac 2 3 	\\
&E^{-R}_1	&	-1	&	 -1	\\%
&e^{-L}_{2,3}	&	-1 	&	 -\frac 1 3 	\\
&e^{-R}_{2}	&	-1 	&	 -1	\\
&\nu^{L}_{2,3}	&	0 	&	 -\frac 1 3 	\\
&N^{0L}_{2,3}	&	0 	&	 -\frac 1 3 	\\
&E^{-L}_4	&	-1 	&	 -\frac 1 3 	\\
&N^{0L}_4	&	0 	&	 -\frac 1 3 	\\
&P^{0L}_4	&	0 	&	 -\frac 1 3 	\\
&N^{0L}_5	&	0 	&	\frac 2 3	\\
\bottomrule
\end{array}
\]
\caption{Fermionic content of the model and associated $U(1)$ charges.}
\label{tab:fermi}
\end{table}

\section{Higgs fields and Yukawa Lagrangian}
\label{app:higgs}

We need to build gauge invariant terms for the coupling between a Higgs field and two fermions, so that we obtain appropriate mass terms after SSB. This constrains possible representations for the scalar fields. Since the fermions transform either as a $3$ or as a $\bar 3$ under $SU(3)_{\textrm L}$, we only have a limited number possibilities~\cite{331:higgs} for a scalar field $\Phi$, which can only be a singlet, a triplet or a sextet~\footnote{We could have also considered antitriplets with opposite charge under $U(1)_X$ with respect to the doublets, and  analogous Yukawa couplings. This would have lead to a doubling of the content in Higgs triplet, but with no further impact on the general discussion outlined here.}.

In the following, we will not analyse the possibility of a singlet scalar. Electromagnetic invariance makes it a scalar under $U(1)_{\textrm X}$. Thus, after the two steps of SSB, its vacuum expectation value will never give rise to a mass term for the gauge bosons or  the charged fermions, and, as indicated before, neutral leptons are outside  the scope of the present work.

\subsection{$SU(3)_\text{L}\times U(1)_X\to SU(2)_\text{L}\times U(1)_{Y}$}

For the first transition $331 \rightarrow 321$, we can have triplet or sextet scalar fields, denoted $\chi$, $\chi^\star$ and $S_1$ respectively. In order to break neither $SU(2)_\text{L}$ nor $U(1)_\text{EM}$ invariances at this stage, the following conditions for vacuum expectation values of the Higgs fields hold
\begin{equation}
	\hat{T}^{1,2,3}\langle \Phi_1 \rangle=\hat Q\langle \Phi_1 \rangle=0, \qquad \Phi_1 \in \{\chi, \chi^\star,  S_1 \}
\label{eq:inv1}
\end{equation}
which sets the v.e.v.s and $U(1)_{X}$ charges of the scalar fields responsible for the first SSB. We have
\begin{equation}\begin{split}
\langle S_1\rangle&=\begin{pmatrix} 	0&0&0\\
			0&0&0\\
			0&0&a_3
\end{pmatrix},\,X=-\frac 2 3
\\
\langle \chi\rangle&=\frac 1 {\sqrt 2} \begin{pmatrix} 0\\0\\u\end{pmatrix},\,X=-\frac 1 3\end{split}
\end{equation}
The Yukawa terms that can be built with the sextet are then of the form
\begin{equation}
	\bar\ell^L_iS_1(\ell^L_j)^c,\quad i,j=2,3,4
\end{equation}
leading only to  Majorana masses for the exotic leptons
$ N^0_{2,3}$, $P^0_4$.

The Yukawa terms built with the triplet and antitriplet contribute to both quarks and lepton mass terms.
The up-quarks  mass terms are of the form
 \begin{equation} \chi^* \bar{Q}_m^L D^R \end{equation}
 where $D^R$ represents both $d_i^R$ and $B_n^R$, with $i=1,2,3$ and $n,m=1,2$. The down-quark mass terms are of the form
	\begin{equation} \bar{Q}_3^L \chi\,  U^R
\end{equation}
	 where $U^R$ represents both $u_i^R$ and $T_3^R$. The equivalent form in the lepton sector is
\begin{equation}
\chi^* \bar{\ell}_1^L L^{-R}
\end{equation} where $L^{-R}$ represents any of $e_{1,2}^{-R}, E_1^{-R}$. The lepton sector also allows combination of $SU(3)_\text{L}$ triplets and antitriplets,  as
\begin{equation}
\epsilon_{ijk}\chi^{*i}\bar\ell^{Lj}_a(\ell_5^L)^{c\,k}
\end{equation} where
the label $a$ can assume values $2,3,4$ and $i,j,k$ are indices referred to $SU(3)_\text{L}$.

\subsection{$ SU(2)_\text{L}\times U(1)_{Y} \to U(1)_\text{EM} $}

The second, electroweak, transition $321 \rightarrow 31$ can involve  two triplets $\eta$ and $\rho$, and sextets, denoted $S_i$.
The electromagnetic gauge invariance still holds after this SSB, which yields the following constraints on the v.e.v.s
\begin{equation}
\hat Q\langle \Phi_2 \rangle=0, \qquad \Phi_2 \in \{\eta, \rho, S_i \}
\label{eq:invApp}
\end{equation}
In order to choose the right alignment for sextet and triplets, we start from the most general ones, impose  a zero charge and verify if we can build Yukawa terms involving these scalar fields and invariant under $U(1)_X$. The
 v.e.v.s of the scalar fields responsible for EWSB are
\begin{equation}\begin{split}
\langle S_b\rangle &=\begin{pmatrix}	b_1&0&b_5\\0&0&0\\b_5&0&b_3\\	\end{pmatrix},\,X=-\frac 2 3
\\
\langle S_c\rangle&=\begin{pmatrix}	0&0&0\\0&c_2&0\\0&0&0\\	\end{pmatrix},\,X=\frac 4 3\\
	\langle \eta\rangle&=\frac{1}{\sqrt 2}\begin{pmatrix}w_1\\0\\w_3\end{pmatrix},\,X=-\frac 1 3
\\
	\langle \rho\rangle&=\frac{1}{\sqrt 2}\begin{pmatrix}0\\v\\0\end{pmatrix},\,X=\frac 2 3\end{split}
\end{equation}
The  $U(1)_X$ invariant terms built with sextets are
\begin{equation}\begin{split}
&\bar\ell^L_iS_b(\ell^L_j)^c,\quad i,j=2,3,4\\
&\bar\ell^L_5S_c(\ell^L_5)^c,\\
&\bar\ell^L_1S^*_c(\ell^L_1)^c.\end{split}
\end{equation}
and for the triplets, we have
\begin{itemize}
	\item for quarks:
	\begin{equation}\begin{split}
	&\bar{Q}_m^L \eta^*D^R,\\
		&\bar{Q}_3^L \eta U^R,\\
		&\bar{Q}^L_3 \rho D^R,\\
		&\bar{Q}^L_m \rho^* U^R;\end{split}
	\end{equation}
	\item for leptons:
	\begin{equation}\begin{split}
&\bar{\ell}_1^L \eta^*L^{-R},\\
		&\bar{\ell}_a^L \rho L^{-R};\\
		&\epsilon_{ijk}\eta^{*i}\bar\ell^{Lj}_a(\ell_5^L)^{c\,k}.\end{split}
\end{equation}	
\end{itemize}
where we have used the same notation of the previous SSB. Therefore, the Yukawa Lagrangian is
\begin{itemize}
	\item for quarks
	\begin{equation}
	\begin{split}
		\mathcal{L}^{q}_Y&=\bigl(\bar Q_m^L\chi^*Y^d_{mi}+\bar Q^L_3\rho y^d_{3i}+\bar Q^L_m \eta^* j^d_{mi}\bigr)D^R_{i}+\\&+\bigl(\bar Q^L_3\chi Y^u_{3j}+\bar Q^L_m\rho^*y^u_{mj}+\bar Q^L_3\eta j^u_{3j}\bigr)U^R_{j},
	\end{split}
		\label{eq:yukq}
	\end{equation}
	where $Y^{d,u}, y^{d,u}, j^{d,u}$ represent the Yukawa couplings introduced respectively for $\chi, \rho$ and $\eta$.
	\item for leptons
\begin{equation}
	\label{eq:yuklep}
	\begin{split}
		\mathcal{L}^{\ell}_Y=&\bigl(\bar\ell^L_1\chi^* Y^{(-)}_{1b}+\bar\ell^L_a\rho f^{(-)}_{ab}+\bar\ell^L_1\eta^* y^{(-)}_{1b}\bigr)L^{-R}_b+\\&+\epsilon_{ijk}(\chi^*)^i(\ell_5^L)^{c\,k}J_a\bar\ell_a^{Lj}+\epsilon_{ijk}(\eta^*)^i(\ell_5^L)^{c\,k}j_{a}\bar\ell_a^{Lj}+\\
&+\bar\ell^L_aS_1(\ell^L_b)^cK_{ab}+\bar\ell^L_aS_b(\ell^L_b)^ck_{ab}+c_5\bar\ell^L_5S_c(\ell^L_5)^c+\\&+c_1\bar\ell^L_1S^*_c(\ell^L_1)^c.
	\end{split}
	\end{equation}
where $Y, y,  K, k, f,  c, J, j$ represent the Yukawa couplings, with $a,b=2,3,4$ and $L_i=e_{1,2}^{-R}, E_1^{-R}$, and where the $i,j,k$ indices are referred to the $SU(3)$ space.
\end{itemize}

\subsection{Quark masses}
\label{app:fermionmasses}

After the two SSBs, the quark mass terms arising from the Yukawa Lagrangian read
	\begin{equation}
	\begin{split}
		\mathcal{L}^{q}_Y\to&\Bigl[\frac u {\sqrt2} \bar B^L_m Y^d_{mi}+\frac v {\sqrt2} \bar d^L_3 y^d_{3i}+\left(\frac{w_1}{\sqrt2} \bar d^L_m+\frac{w_2}{\sqrt2}\bar B^L_m\right)j^d_{mi}\Bigr]D^R_i+\\&+\Bigl[\frac u {\sqrt 2}\bar T^L_3 Y^u_{3i}-\frac v {\sqrt2} \bar u^L_{m} y^u_{mi}+\left(\frac{w_1}{\sqrt2}\bar u^L_3+\frac{w_2}{\sqrt2}\bar T^L_3\right)j^u_{3i}\Bigr]U^R_i.
	\end{split}
		\label{eq:massq}
	\end{equation}
It is possible to rewrite these mass terms in the form of a matrix product with the flavour vectors $D, U$, introduced in Eq. (\ref{eq:flavvec}) as
\begin{equation}
	M_q=\bar D_L M_d D_R + \bar U_L M_u U_R
\end{equation}
where\begin{widetext}
\begin{equation}\begin{split}
M_u=&\frac 1 {\sqrt2}\begin{pmatrix}
	-y^u_{11}v&-y^u_{12}v&-y^u_{13}v&-y^u_{14}v\\
	-y^u_{21}v&-y^u_{22}v&-y^u_{23}v&-y^u_{24}v\\
	j^u_{31}w_1&j^u_{32}w_1&j^u_{33}w_1&j^u_{34}w_1\\
	j^u_{31}w_2+Y^u_{31}u&j^u_{32}w_2+Y^u_{32}u&j^u_{33}w_2+Y^u_{33}u&j^u_{34}w_2+Y^u_{34}u\\
\end{pmatrix}\\
M_d=&\frac 1 {\sqrt2}\begin{pmatrix}
	j^d_{11}w_1&j^d_{12}w_1&j^d_{13}w_1&j^d_{14}w_1&j^d_{15}w_1\\
	j^d_{21}w_1&j^d_{22}w_1&j^d_{23}w_1&j^d_{24}w_1&j^d_{25}w_1\\
	y^d_{31}v&y^d_{32}v&y^d_{33}v&y^d_{34}v&y^d_{35}v\\
	j^d_{11}w_2+Y^d_{11}u&j^d_{12}w_2+Y^d_{12}u&j^d_{13}w_2+Y^d_{13}u&j^d_{14}w_2+Y^d_{14}u&j^d_{15}w_2+Y^d_{15}u\\
	j^d_{21}w_2+Y^d_{21}u&j^d_{22}w_2+Y^d_{22}u&j^d_{23}w_2+Y^d_{23}u&j^d_{24}w_2+Y^d_{24}u&j^d_{25}w_2+Y^d_{25}u\\
\end{pmatrix}\end{split}
\end{equation}\end{widetext}

The diagonalisation in the limit $v=w_1=w_2=0$ (before the EWSB) shows that the number of quarks that remain massless after the $SU(3)_{\textrm L}$ SSB is three for up-type and three for down-type quarks (for a given colour). This is exactly equal to the number of SM particles, meaning that all the new exotic particles acquire a mass of the scale $\Lambda_{NP}$ of the $SU(3)_{\textrm L}$ SSB. This feature of the model is required if we want to justify why such particles have not yet been observed at the electroweak scale.

\subsection{Charged lepton masses}
\label{app:fermionmasses1}

In our model, we have identified the charged elements of $\ell_5$ with the charge conjugated right handed components of particles already introduced in other generations; to be more precise, we have set
\begin {equation}
\ell^L_5=\begin{pmatrix}E^{+L}_5 \\ N^{0L}_5 \\ F^{+L}_5 \end{pmatrix} \rightarrow \begin{pmatrix}\bigl(E^{-R}_4\bigr)^c\\ N^{0L}_5 \\ \bigl(e^{-R}_3\bigr)^c\end{pmatrix}.
\label{lhBappend}
\end{equation}

 Apart from limiting the number of additional degrees of freedom, the main reason of this identification is not clear until we consider the charged exotic masses.

Without such identification, the introduction of the right-handed degrees of freedom of the charged leptons appearing in the fifth generation implies the additional  Yukawa term
\begin{equation}
	\label{eq:yukl1}
	\begin{split}
		\mathcal{L}^{\ell}_Y\supset\bar\ell^L_5\bigl(\chi Y^{(+)}_{5k}+\eta y^{(+)}_{5k}\bigr)P^{+R}_k.
	\end{split}
	\end{equation}
	where $P^{+R}$ represents the right handed components of the positively charged elements $E^+_5$, $F^+_5$ of $\ell_5$. Furthermore, the vector $L_a$ in~\eqref{eq:yuklep} stands now for $L_i=e_{1,2,3}^{-R}, E_{1,4}^{-R}$. Introducing the flavour vector for negatively charged leptons
\begin{equation}
	\begin{pmatrix}
		e_1&
		e_2&
		e_3&
		E_1&
		E_4&
		E_5^c&
		F_5^c
	\end{pmatrix}^T
\end{equation}
after the first SSB we get the following mass matrix
\begin{equation}
	\begin{pmatrix}
	0&0&0&0&0&0&0\\
	0&0&0&0&0&-\frac{J_{15}u^*}{\sqrt 2}&0\\
	0&0&0&0&0&-\frac{J_{25}u^*}{\sqrt 2}&0\\
	\frac{Y^{(-)}_{11}u^*}{\sqrt 2}&\frac{Y^{(-)}_{12}u^*}{\sqrt 2}&\frac{Y^{(-)}_{13}u^*}{\sqrt 2}&\frac{Y^{(-)}_{14}u^*}{\sqrt 2}&\frac{Y^{(-)}_{15}u^*}{\sqrt 2}&0&0\\
	0&0&0&0&0&-\frac{J_{35}u^*}{\sqrt 2}&0\\
	0&0&0&0&0&0&0\\
	0&0&0&0&0&\frac{Y^{(+)}_{5E}u}{\sqrt 2}&\frac{Y^{(+)}_{5F}u}{\sqrt 2}\\
	\end{pmatrix}
\end{equation}
One can check that the degeneracy of the 0 eigenvalue of this matrix is greater than 3, implying that out of all the charged leptons, not just the ones to be identified with the SM ones acquire mass at the EW scale.

As indicated in Sec.~\ref{sec:fieldsandrepr}, we  avoid the presence of charged exotic particles with masses of the order of the EW scale, which have not been observed phenomenologically, through the identification of the charged elements of $\ell_5$ with the charge conjugates of the right-handed components of particles already introduced for other generations.
With this assumption, the mass matrix of charged leptons originating after the two stages of SSB becomes~\cite{sher:331}
\begin{widetext}\begin{equation}
M_e=\frac 1 {\sqrt2}	\begin{pmatrix}
		y_{e_1}w_1&y_{e_2}w_1&0&y_{E_1}w_1&0\\
		k_{2e_1}v&k_{2e_2}v&j_{e_2}w_1&k_{2E_1}v&-J_{e_2}u-j_{e_2}w_2\\
		k_{3e_1}v&k_{3e_2}v&j_{e_3}w_1&k_{3E_1}v&-J_{e_3}u-j_{e_3}w_2\\
		Y_{e_1}u+y_{e_1}w_2&Y_{e_2}u+y_{e_2}w_2&0&Y_{E_1}u+y_{E_1}w_2&0\\
		k_{4e_1}v&k_{4e_2}v&j_{E_4}w_1&k_{4E_1}v&-J_{E_4}u-j_{E_4}w_2\\
	\end{pmatrix}
\end{equation}\end{widetext}

The diagonalisation in the limit $v=w_1=w_2=0$ (before the EWSB) shows that the number of leptons that remain massless after the $SU(3)_{\textrm L}$ SSB is three. This is exactly equal to the number of SM particles, meaning that all the new exotic particles acquire a mass of the scale $\Lambda_{NP}$ of the $SU(3)_{\textrm L}$ SSB. This feature of the model is required if we want to justify why such particles have not yet been observed at the electroweak scale.

\section{Anomaly cancellation} \label{app:anomalycancel}

Particularly stringent constraints for 331 model building arise from requiring that the theory is free from quantum anomalies. We list here the relations among the fermion charges that need to be satisfied. We denote with $Q$ the quark left-handed generations, $q$ the corresponding singlets, $\ell$ the leptonic multiplets and $s$ the corresponding singlets. Imposing the vanishing of the triangular anomaly coupling to the different gauge bosons of the theory leads to~\cite{331:anomaly}
\begin{equation}
[SU(3)_c]^2\otimes U(1)_X \Rightarrow 3\sum_{Q} X^L_Q - \sum_q X_q^R=0\label{eq:anom}\end{equation}
\begin{equation}
[SU(3)_L]^3 \Rightarrow \; \parbox{15em}{equal number of $3$ and $\bar 3$ fermionic representations}\label{eq:anom2}
\end{equation}\begin{equation}
[SU(3)_L]^2\otimes U(1)_X \Rightarrow 3\sum_{Q} X^L_Q + \sum_\ell X_\ell^L=0\end{equation}
\begin{equation}\begin{split}[\text{Grav}]^2\otimes U(1)_X \Rightarrow &9\sum_{Q} X^L_Q + 3 \sum_\ell X_\ell^L+\\&- 3 \sum_q X_q^R- \sum_s X_s^R=0\end{split}\end{equation}\begin{equation}\begin{split}
[U(1)_X]^3 \Rightarrow &9\sum_{Q} (X^L_Q)^3 + 3 \sum_\ell (X_\ell^L)^3+\\&- 3 \sum_q (X_q^R)^3- \sum_s (X_s^R)^3=0\label{eq:anom3}
\end{split}\end{equation}
It is clear from Eq.~\eqref{eq:anom2} that we cannot generate LFUV couplings for the gauge bosons unless we introduce additional lepton families. Indeed, if we call $N_Q$ ($N_{\bar{Q}}$)  the number of quark generations transforming as a $3$ ($\overline 3$), with similar notation for the leptons $\ell$, the anomaly cancellation in Eq.~\eqref{eq:anom2} yields
\begin{equation}
	3 N_Q - 3 N_{\overline Q} +N_\ell-N_{\overline\ell}=0\,.
\end{equation}
Restricting to just three generations of quarks $N_Q+N_{\bar{Q}}=3$, we see that
one has several possibilities.
If we assume that all three quark families transform in the same way, one needs at least 9 lepton generations (3 SM leptonics and  6 exotic ones), which would then transform all in the same opposite way to get the appropriate anomaly cancellation. Since all leptons transform in the same way, there is no possibility to generate different couplings between the leptons and the gauge bosons, and thus no LFUV can arise from these couplings.

The situation changes  if one of the quark families transforms differently compared to the others. Indeed, if we assume only two quark families to transform as a $\overline 3$, we obtain
\begin{equation}
	N_\ell-N_{\overline\ell}=3
	\label{leptonfamilies}
\end{equation}
Assuming three lepton generations implies that $N_\ell=3, N_{\overline\ell}=0$. In this minimal model, often considered in the literature, there is no possibility to generate LFUV from the identical couplings of the gauge bosons to all lepton families.
We can increase the number of lepton generations. Assuming four generations, i.e., $N_\ell+N_{\overline\ell}=4$, yields no integer solutions for Eq.~\eqref{leptonfamilies}. The next possibility is $N_\ell+N_{\overline\ell}=5$ lepton families, so that $N_\ell=4, N_{\overline\ell}=1$, which provides LFUV in the gauge couplings to leptons~\cite{331:anomaly}.
This is the
non-minimal choice that we adopt.

\onecolumngrid
\section{Currents}

\label{app:curr}

We provide the expression of the couplings of the gauge bosons with the fermions, the latter being expressed in the interaction eigenbasis.
\subsection{Charged currents}
For the non-SM charged gauge boson $V^\pm$ we get
\scriptsize
\begin{equation}
\begin{split}
	&\mathcal L_V=\frac g {\sqrt2}V^-_\mu\Bigl\{\bar D^L\gamma^\mu \begin{pmatrix}0&0&0&0\\0&0&0&0\\0&0&0&1\\1&0&0&0\\0&1&0&0\end{pmatrix}U^L+\bar N^L\gamma^\mu\begin{pmatrix} 0&0&0&0&0\\0&0&0&0&0\\0&0&0&0&0\\0&0&0&0&0\\0&0&0&0&0\\0&0&0&0&0\\0&0&1&0&0\\0&0&0&0&0\end{pmatrix}(f^{-R})^c+\bar f^{-L}\gamma^\mu\begin{pmatrix}0&0&0&0&0&0&0&0\\0&0&0&1&0&0&0&0\\0&0&0&0&1&0&0&0\\1&0&0&0&0&0&0&0\\0&0&0&0&0&0&0&1\end{pmatrix}N^L\Bigr\}+\\
	&+\frac g {\sqrt2}V^+_\mu\Bigl\{\bar U^L\gamma^\mu \begin{pmatrix}0&0&0&1&0\\0&0&0&0&1\\0&0&0&0&0\\0&0&1&0&0\end{pmatrix}D^L+\bar N^L\gamma^\mu\begin{pmatrix} 0&0&0&1&0\\0&0&0&0&0\\0&0&0&0&0\\0&1&0&0&0\\0&0&1&0&0\\0&0&0&0&0\\0&0&0&0&0\\0&0&0&0&1\end{pmatrix}f^{-L}+\overline{ (f^{-R})^c}\gamma^\mu\begin{pmatrix}0&0&0&0&0&0&0&0\\0&0&0&0&0&0&0&0\\0&0&0&0&0&0&1&0\\0&0&0&0&0&0&0&0\\0&0&0&0&0&0&0&0\\\end{pmatrix}N^L\Bigr\}.
\end{split}
\end{equation}
\normalsize
For the SM charged gauge bosons $W^\pm$ we get
\scriptsize
\begin{equation}
\begin{split}
	&\mathcal L_W=\frac g {\sqrt2}W^-_\mu\Bigl\{\bar D^L\gamma^\mu \begin{pmatrix}1&0&0&0\\0&1&0&0\\0&0&1&0\\0&0&0&0\\0&0&0&0\end{pmatrix}U^L+\bar N^L\gamma^\mu\begin{pmatrix} 0&0&0&0&0\\0&0&0&0&0\\0&0&0&0&0\\0&0&0&0&0\\0&0&0&0&0\\0&0&0&0&0\\0&0&0&0&1\\0&0&0&0&0\end{pmatrix}(f^{-R})^c+\bar f^{-L}\gamma^\mu\begin{pmatrix}1&0&0&0&0&0&0&0\\0&1&0&0&0&0&0&0\\0&0&1&0&0&0&0&0\\0&0&0&0&0&0&0&0\\0&0&0&0&0&1&0&0\end{pmatrix}N^L\Bigr\}\\
	&+\frac g {\sqrt2}W^+_\mu\Bigl\{\bar U^L\gamma^\mu \begin{pmatrix}1&0&0&0&0\\0&1&0&0&0\\0&0&1&0&0\\0&0&0&0&0\end{pmatrix}D^L+\bar N^L\gamma^\mu\begin{pmatrix} 1&0&0&0&0\\0&1&0&0&0\\0&0&1&0&0\\0&0&0&0&0\\0&0&0&0&0\\0&0&0&0&1\\0&0&0&0&0\\0&0&0&0&0\end{pmatrix}f^{-L}+\overline {(f^{-R})^c}\gamma^\mu\begin{pmatrix}0&0&0&0&0&0&0&0\\0&0&0&0&0&0&0&0\\0&0&0&0&0&0&0&0\\0&0&0&0&0&0&0&0\\0&0&0&0&0&0&1&0\\\end{pmatrix}N^L\Bigr\}
\end{split}
\label{eq:W}
\end{equation}
\normalsize
In the previous relations  the flavour vectors of charged fields $D$, $U$ and $f^-$ have been introduced in Sect.\ref{eq:flavvec}, and the neutral flavour vector is defined as
$N \equiv (\nu_1, \nu_2, \nu_3, N_2^0, N_3^0, N_4^0, N_5^0, P_4^0)$.

\subsection{Neutral currents}

First we provide the interactions with the non-SM neutral gauge bosons $W^{4,5}, Z'$
\scriptsize
\begin{equation}
\begin{split}
	\mathcal L_4&=\frac g {2}W^4_\mu\Bigl\{\bar U^L \gamma^\mu\begin{pmatrix}0&0&0&0\\0&0&0&0\\0&0&0&1\\0&0&1&0\end{pmatrix}U^L-\bar D^L \gamma^\mu\begin{pmatrix}0&0&0&1&0\\0&0&0&0&1\\0&0&0&0&0\\1&0&0&0&0\\0&1&0&0&0\end{pmatrix} D^L+\\
	-&\bar f^{-L}\gamma^\mu\begin{pmatrix}0&0&0&1&0\\0&0&0&0&0\\0&0&0&0&0\\1&0&0&0&0\\0&0&0&0&0\end{pmatrix}f^{-L}
+\bar N^L\gamma^\mu\begin{pmatrix}0&0&0&0&0&0&0&0\\0&0&0&1&0&0&0&0\\0&0&0&0&1&0&0&0\\0&1&0&0&0&0&0&0\\0&0&1&0&0&0&0&0\\0&0&0&0&0&0&0&1\\0&0&0&0&0&0&0&0\\0&0&0&0&0&1&0&0\end{pmatrix}N^L-\bar f^{-R}\gamma^\mu\begin{pmatrix}0&0&0&0&0\\0&0&0&0&0\\0&0&0&0&1\\0&0&0&0&0\\0&0&1&0&0\end{pmatrix} f^{-R}\Bigr\}
\end{split}
\label{eq:W4}
\end{equation}

\begin{equation}
\begin{split}
	\mathcal L_5&=\frac ig {2}W^5_\mu\Bigl\{\bar U^L \gamma^\mu\begin{pmatrix}0&0&0&0\\0&0&0&0\\0&0&0&-1\\0&0&1&0\end{pmatrix}U^L+\bar D^L \gamma^\mu\begin{pmatrix}0&0&0&-1&0\\0&0&0&0&-1\\0&0&0&0&0\\1&0&0&0&0\\0&1&0&0&0\end{pmatrix} D^L+\\
	+&\bar f^{-L}\gamma^\mu\begin{pmatrix}0&0&0&-1&0\\0&0&0&0&0\\0&0&0&0&0\\1&0&0&0&0\\0&0&0&0&0\end{pmatrix}f^{-L}
+\bar N^L\gamma^\mu \begin{pmatrix}0&0&0&0&0&0&0&0\\0&0&0&-1&0&0&0&0\\0&0&0&0&-1&0&0&0\\0&1&0&0&0&0&0&0\\0&0&1&0&0&0&0&0\\0&0&0&0&0&0&0&-1\\0&0&0&0&0&0&0&0\\0&0&0&0&0&1&0&0\end{pmatrix}N^L-\bar f^{-R}\gamma^\mu\begin{pmatrix}0&0&0&0&0\\0&0&0&0&0\\0&0&0&0&1\\0&0&0&0&0\\0&0&-1&0&0\end{pmatrix} f^{-R}\Bigr\}
\end{split}
\label{eq:W5}
\end{equation}

\begin{equation}
\label{eq:Z'}
\centering
\begin{split}
	&\mathcal L_{Z'}=\frac {\cos\theta_{331}}{g_X}Z'_\mu\Bigl\{\bar U^L\gamma^\mu \begin{pmatrix}-\sqrt{\frac 3 2}g^2&0&0&0\\0&-\sqrt{\frac 3 2}g^2&0&0\\0&0&\frac{9g^2+g_X^2}{3\sqrt6}&0\\0&0&0&\frac{-18g^2+g_X^2}{3\sqrt6}\end{pmatrix}U^L+\frac{\sqrt2g_X^2}{3\sqrt3}\bar U^R\gamma^\mu U^R+\\
	+\bar D^L\gamma^\mu&\begin{pmatrix}-\sqrt{\frac 3 2}g^2&0&0&0&0\\0&-\sqrt{\frac 3 2}g^2&0&0&0\\0&0&\frac{9g^2+g_X^2}{3\sqrt6}&0&0\\0&0&0&\sqrt6g^2&0\\0&0&0&0&\sqrt6g^2\\\end{pmatrix} D^L-\frac{g_X^2}{3\sqrt6}\bar D^R\gamma^\mu D^R+\\
	&-\bar f^{-L}\gamma^\mu\begin{pmatrix}\frac{9g^2+2g_X^2}{3\sqrt6}&0&0&0&0\\0&\frac{-9g^2+g_X^2}{3\sqrt6}&0&0&0\\0&0&\frac{-9g^2+g_X^2}{3\sqrt6}&0&0\\0&0&0&\frac{2(-9g^2+g_X^2)}{3\sqrt6}&0\\0&0&0&0&\frac{-9g^2+g_X^2}{3\sqrt6}\end{pmatrix}f^{-L}+\\
	&+\bar f^{-R}\gamma^\mu \begin{pmatrix}\frac{g_X^2}{\sqrt6}&0&0&0&0\\0&\frac{g_X^2}{\sqrt6}&0&0&0\\0&0&\frac{\sqrt2(9g^2-g_X^2)}{3\sqrt3}&0&0\\0&0&0&\frac{g_X^2}{\sqrt6}&0\\0&0&0&0&-\frac{\sqrt2(9g^2+2g_X^2)}{3\sqrt3}\end{pmatrix}f^{-R}+\\
	+\bar N^L\gamma^\mu&\begin{pmatrix}-\frac{9g^2+2g_X^2}{3\sqrt6}&0&0&0&0&0&0&0\\0&-\frac{-9g^2+g_X^2}{3\sqrt6}&0&0&0&0&0&0\\0&0&-\frac{-9g^2+g_X^2}{3\sqrt6}&0&0&0&0&0\\0&0&0&-\frac{18g^2+g_X^2}{3\sqrt6}&0&0&0&0\\0&0&0&0&-\frac{18g^2+g_X^2}{3\sqrt6}&0&0&0\\0&0&0&0&0&-\frac{-9g^2+g_X^2}{3\sqrt6}&0&0\\0&0&0&0&0&0&\frac{9g^2+2g_X^2}{3\sqrt6}&0\\0&0&0&0&0&0&0&-\frac{18g^2+g_X^2}{3\sqrt6}\end{pmatrix}N^L\Bigr\}
\end{split}
\end{equation}
\normalsize
Moving to the SM neutral gauge bosons $Z,A$, we have
\scriptsize
\begin{equation}
\label{eq:Z}
\begin{split}
	\mathcal L_Z&=\cos\theta_W gZ_\mu\Bigl\{\bar U^L\gamma^\mu \begin{pmatrix}\frac {1-\cos^2\theta_{331}}2&0&0&0 \\0&\frac {1-\cos^2\theta_{331}}2&0&0 \\0&0&\frac {1-\cos^2\theta_{331}}2&0\\0&0&0&-2\cos^2\theta_{331}\end{pmatrix}U^L-2\cos^2\theta_{331}\bar U^{R}\gamma^\mu U^R+\\
	&+\bar D^L\gamma^\mu \begin{pmatrix}-\frac {1+\cos^2\theta_{331}}2&0&0&0 \\0&-\frac {1+\cos^2\theta_{331}}2&0&0&0 \\0&0&-\frac {1+\cos^2\theta_{331}}2&0&0\\0&0&0&\cos^2\theta_{331}&0\\0&0&0&0&\cos^2\theta_{331} \end{pmatrix}D^L+\cos^2\theta_{331}\bar D^{R}\gamma^\mu D^R+\\
	&+\bar f^{-L}\gamma^\mu\begin{pmatrix} \frac{-1+3\cos^2\theta_{331}}2&0&0&0&0\\0&\frac{-1+3\cos^2\theta_{331}}2&0&0\\0&0&\frac{-1+3\cos^2\theta_{331}}2&0&0\\0&0&0&3\cos^2\theta_{331}&0\\0&0&0&0&\frac{-1+3\cos^2\theta_{331}}2 \end{pmatrix}f^{-L}+\\
	& +\bar f^{-R}\gamma^\mu\begin{pmatrix}3\cos^2\theta_{331}&0&0&0&0\\0&3\cos^2\theta_{331}&0&0&0\\0&0&3\cos^2\theta_{331}&0&0\\0&0&0&3\cos^2\theta_{331}&0\\0&0&0&0&\frac{-1+3\cos^2\theta_{331}}{2}\end{pmatrix}f^{-R}+\\
	&+\frac{1+3\cos^2\theta_{331}}2\bar N^L\gamma^\mu\begin{pmatrix}1&0&0&0&0&0&0&0\\0&1&0&0&0&0&0&0\\0&0&1&0&0&0&0&0\\0&0&0&0&0&0&0&0\\0&0&0&0&0&0&0&0\\0&0&0&0&0&1&0&0\\0&0&0&0&0&0&-1&0\\0&0&0&0&0&0&0&0\end{pmatrix}N^L\Bigr\}
\end{split}
\end{equation}
%
\begin{equation}
	\mathcal L_A=\sqrt3\cos\theta_{331}\cos\theta_W gA_\mu\Bigl\{-\frac 2 3\bar U\gamma^\mu U+\frac 1 3\bar D\gamma^\mu D+\bar f^-\gamma^\mu f^-\Bigr\}
\label{eq:A}
\end{equation}
\normalsize

\section{Four-fermion operators involving light charged fermions mediated by neutral currents
\label{sec:ZZprimecouplings}}

 We want to determine the contributions for four-fermion operators up to and including $O(\epsilon^2)$ in the effective Hamiltonian involving  light charged fermions and mediated by neutral gauge bosons.
It turns out that the only relevant couplings are the ones between $Z$ and $Z^\prime$ to light charged fermions.

In the case of the $Z$ boson, it means that we have to determine the $O(\epsilon^2)$ corrections to  the SM couplings $O(\epsilon^0)$.
For each chirality of each fermion type $U,D,f,N$, it proves useful to split the $Z$ coupling Eq.~(\ref{eq:Z}) between a contribution proportional to the identity that is the only contribution for SM fermions and a contribution only for exotic fermions, e.g.:
\scriptsize
\begin{equation}
\begin{split}
	\mathcal L_Z&=\cos\theta_W gZ_\mu\left\{\frac{1-\cos^2\theta_{331}}{2} \bar U^L\gamma^\mu
	 \begin{pmatrix}1&0&0&0 \\0&1&0&0 \\0&0&1&0\\0&0&0&1\end{pmatrix}U^L
	   -\frac{1+3\cos^2\theta_{331}}{2}\bar U^L\gamma^\mu \begin{pmatrix}0&0&0&0 \\0&0&0&0 \\0&0&0&0\\0&0&0&1\end{pmatrix}U^L +\ldots\right\}
\end{split}
\end{equation}
\normalsize
which can be expressed in terms of mass eigenstates using the rotations $V$ and $W$ defined in Eq.~(\ref{eq:VWrotations}). The first term, proportional to identity, is unaffected by the rotations. The second term can induce couplings to SM through mixing to exotic fermions: this cannot come from $O(\epsilon^0)$ $V$ and $W$ as they are block-diagonal, connecting only SM fermions among themselves and exotic fermions among themselves, but it can occur from their $O(\epsilon^1)$ contributions, denoted $\hat{V}$ and $\hat{W}$, which connect SM and exotic fermions. At $O(\epsilon^2)$ in the couplings, one thus obtains the couplings for the $Z$ meson to SM fermions in the mass eigenbasis:
\begin{equation}
	\mathcal L_Z\supset  \cos\theta_W gZ_\mu \sum_{\psi=u,d,f^-} \sum_{X=L,R} \sum_{k,l=1,2,3}
	(\alpha+\beta)^{\psi^X}_{kl} \bar{\psi}^X_k \gamma^\mu \psi^X_l
\end{equation}
where $\alpha$ and $\beta$ correspond to SM $O(\epsilon^0)$ and NP $O(\epsilon^2)$ couplings respectively. Their values are collected in Tab.~\ref{tab:ZZprimecouplings}.

\begin{table}[t]
$$
\begin{array}{cccc}
\toprule
 & \alpha_{kl} & \beta_{kl} & \gamma_{kl} \\
\midrule
u^L & \frac{1}{2}(1-\cos^2\theta_{331})\delta_{kl} &  -\frac{1}{2}(1+3\cos^2\theta_{331})\hat{V}^{(u)*}_{4k}\hat{V}^{(u)}_{4l}  & -\frac{1}{2}(1-\cos^2\theta_{331})\delta_{kl}+V^{(u)*}_{3k}V^{(u)}_{3l}\\
u^R & -2\cos^2\theta_{331}\delta_{kl} & 0 & 2\cos^2\theta_{331}\delta_{kl}\\
d^L & -\frac{1}{2}(1+\cos^2\theta_{331})\delta_{kl} &  \frac{1}{2}(1+3\cos^2\theta_{331})(\hat{V}^{(d)*}_{4k}\hat{V}^{(d)}_{4l}+\hat{V}^{(d)*}_{5k}\hat{V}^{(d)}_{5l})
& -\frac{1}{2}(1-\cos^2\theta_{331})\delta_{kl}+V^{(d)*}_{3k}V^{(d)}_{3l}\\
d^R &  \cos^2\theta_{331}\delta_{kl} & 0 & -\cos^2\theta_{331}\delta_{kl}\\
f^{-L} &\frac{1}{2}(-1+3\cos^2\theta_{331})\delta_{kl} &  \frac{1}{2}(1+3\cos^2\theta_{331})\hat{V}^{(e)*}_{4k}\hat{V}^{(e)}_{4l} & \frac{1}{2}(1-3\cos^2\theta_{331})\delta_{kl}-V^{(e)*}_{1k}V^{(e)}_{1l}\\
f^{-R} & 3\cos^2\theta_{331}\delta_{kl} &  -\frac{1}{2}(1+3\cos^2\theta_{331})\hat{W}^{(e)*}_{5k}\hat{W}^{(e)}_{5l}
& 3\cos^2\theta_{331}\delta_{kl}+(1-6\cos^2\theta_{331})W^{(e)*}_{3k}W^{(e)}_{3l}\\
\bottomrule
\end{array}
$$
\caption{$Z$ and $Z'$ couplings to light charged fermions up to $O(\epsilon^2)$. $V$ and $W$ unitary matrices can be considered at $O(\epsilon^0)$ only, whereas $\hat{V}$ and $\hat{W}$ denote their $O(\epsilon^1)$ components.\label{tab:ZZprimecouplings}}
\end{table}

A similar analysis can be carried out for the interaction with $Z^\prime$ starting from Eq.~(\ref{eq:Z'}). The propagation of the heavy $Z'$ boson provides already a $O(\epsilon^2)$ suppression for the effective four-fermion operators, so we have only to consider the $O(\epsilon^0)$ couplings of the $Z'$ to light charged fermions. We can determine these couplings by splitting Eq.~(\ref{eq:Z'}) into
a term proportional to the identity in flavour space and a term that depends on the generation and we reexpress all the fermion fields in the mass eigenbasis using Eq.~(\ref{eq:VWrotations}). We have only to consider the $O(\epsilon^0)$ part of these rotations, which connect only SM flavours among themselves and exotic flavours among themselves. As we are only interested in the coupling of the $Z'$ to light charged fermions, we can restrict the analysis to the SM sector, leading to the following structure of couplings:
\begin{equation}
	\mathcal L_{Z'}\supset  \frac{1}{3\sqrt{6}} \frac{g_X}{\cos\theta_{331}} Z'_\mu \sum_{\psi=u,d,f^-} \sum_{X=L,R} \sum_{k,l=1,2,3}
	\gamma^{\psi^X}_{kl} \bar{\psi}^X_k \gamma^\mu \psi^X_l
\end{equation}
where $\gamma$ correspond to NP $O(\epsilon^0)$ couplings. Their values are collected in Tab.~\ref{tab:ZZprimecouplings}.

There are no further contributions to be considered from the other neutral gauge bosons for neutral currents. Indeed,
for the photon $A$, we see from Eq.~\eqref{eq:A} that the interaction with down-type quarks is proportional to the identity matrix in flavour space, so that there are no FCNC from the photon interaction.
Concerning $W^{4,5}$, we see from Eqs.~\eqref{eq:W4} and~\eqref{eq:W5} that these gauge bosons always couple
a SM particle with an exotic one in the interaction basis, which occurs only at order $O(\epsilon)$. Furthermore, the process is mediated by a heavy gauge boson, adding a further $O(\epsilon^2)$ suppression. Therefore the $W^{4,5}$ contributions to the process are of order $O(\epsilon^3)$ and  can be neglected compared to the $O(\epsilon^2)$ NP contributions from $Z$ and $Z'$ gauge bosons.

The $O(\epsilon^2)$  NP corrections induced to the effective Hamiltonian will be of the form
\begin{equation}
\begin{split}
{\mathcal H}_{eff}&\supset
4\sqrt{2}G_F\cos^4\theta_W
\sum_{X,Y=L,R} \sum_{\psi,\Psi=u,d,f^-}
\left[\alpha_{ij}\beta_{kl}+\beta_{ij}\alpha_{kl}+
\frac{1}{4\cos^2\theta_W-1}\frac{M_Z^2}{M_{Z'}^2}\gamma_{ij}\gamma_{kl}\right]
    (\bar\psi^X_i\gamma^\mu\psi^X_j)(\bar\Psi^Y_k\gamma_\mu\Psi^Y_l)\\
    &\qquad\qquad =4\sqrt{2}G_F\cos^4\theta_W
\sum_{X,Y=L,R} \sum_{\psi,\Psi=u,d,f^-}
\Delta[\psi^X_i,\psi^X_j,\Psi^Y_k,\Psi^Y_l]    (\bar\psi^X_i\gamma^\mu\psi^X_j)(\bar\Psi^Y_k\gamma_\mu\Psi^Y_l)
    \end{split}\label{eq:Delta}
\end{equation}
We see that the couplings $\Delta[\psi^X_i,\psi^X_j,\Psi^Y_k,\Psi^Y_l]$ are of $O(\epsilon^2)$ and combine $Z$ and $Z'$ couplings.

\twocolumngrid


\bibliographystyle{ieeetr}
\nocite{*}
\bibliography{331}

\begin{thebibliography}{10}

\bibitem{Ricciardi:2016jjb}
G.~Ricciardi, ``{Tensions in the flavour sector},'' {\em EPJ Web Conf.},
  vol.~137, p.~06022, 2017.

\bibitem{Capdevila:2017ert}
B.~Capdevila, S.~Descotes-Genon, L.~Hofer, and J.~Matias, ``{Hadronic
  uncertainties in $B \to K^* \mu^+ \mu^-$: a state-of-the-art analysis},''
  {\em JHEP}, vol.~04, p.~016, 2017.

\bibitem{Koppenburg:2017mad}
S.~Descotes-Genon and P.~Koppenburg, ``{The CKM Parameters},'' {\em Ann. Rev.
  Nucl. Part. Sci.}, vol.~67, pp.~97--127, 2017.

\bibitem{Ricciardi:2016pmh}
G.~Ricciardi, ``{Semileptonic and leptonic $B$ decays, circa 2016},'' {\em Mod.
  Phys. Lett.}, vol.~A32, no.~05, p.~1730005, 2017.

\bibitem{Rk:LHCb}
R.~Aaij {\em et~al.}, ``{Test of lepton universality using $B^{+}\rightarrow
  K^{+}\ell^{+}\ell^{-}$ decays},'' {\em Phys. Rev. Lett.}, vol.~113,
  p.~151601, 2014.

\bibitem{Rkst:LHCb}
R.~Aaij {\em et~al.}, ``{Test of lepton universality with $B^{0} \rightarrow
  K^{*0}\ell^{+}\ell^{-}$ decays},'' {\em JHEP}, vol.~08, p.~055, 2017.

\bibitem{RD:HFAG}
Y.~Amhis {\em et~al.}, ``{Averages of $b$-hadron, $c$-hadron, and $\tau$-lepton
  properties as of summer 2016},'' {\em Eur. Phys. J.}, vol.~C77, no.~12,
  p.~895, 2017.

\bibitem{Bordone:2016gaq}
M.~Bordone, G.~Isidori, and A.~Pattori, ``{On the Standard Model predictions
  for $R_K$ and $R_{K^*}$},'' {\em Eur. Phys. J.}, vol.~C76, no.~8, p.~440,
  2016.

\bibitem{RDs:th}
H.~Na, C.~M. Bouchard, G.~P. Lepage, C.~Monahan, and J.~Shigemitsu, ``{$B
  \rightarrow D l \nu$ form factors at nonzero recoil and extraction of
  $|V_{cb}|$},'' {\em Phys. Rev.}, vol.~D92, no.~5, p.~054510, 2015.
\newblock [Erratum: Phys. Rev.D93,no.11,119906(2016)].

\bibitem{Fajfer:2012vx}
S.~Fajfer, J.~F. Kamenik, and I.~Nisandzic, ``{On the $B \to D^* \tau \bar
  \nu_{\tau}$ Sensitivity to New Physics},'' {\em Phys. Rev.}, vol.~D85,
  p.~094025, 2012.

\bibitem{RD:BaBar1}
J.~P. Lees {\em et~al.}, ``{Evidence for an excess of $\bar{B} \to D^{(*)}
  \tau^-\bar{\nu}_\tau$ decays},'' {\em Phys. Rev. Lett.}, vol.~109, p.~101802,
  2012.

\bibitem{RD:BaBar2}
J.~P. Lees {\em et~al.}, ``{Measurement of an Excess of $\bar{B} \to
  D^{(*)}\tau^- \bar{\nu}_\tau$ Decays and Implications for Charged Higgs
  Bosons},'' {\em Phys. Rev.}, vol.~D88, no.~7, p.~072012, 2013.

\bibitem{RD:LHCb}
R.~Aaij {\em et~al.}, ``{Measurement of the ratio of branching fractions
  $\mathcal{B}(\bar{B}^0 \to
  D^{*+}\tau^{-}\bar{\nu}_{\tau})/\mathcal{B}(\bar{B}^0 \to
  D^{*+}\mu^{-}\bar{\nu}_{\mu})$},'' {\em Phys. Rev. Lett.}, vol.~115, no.~11,
  p.~111803, 2015.
\newblock [Erratum: Phys. Rev. Lett.115,no.15,159901(2015)].

\bibitem{RD:Belle}
M.~Huschle {\em et~al.}, ``{Measurement of the branching ratio of $\bar{B} \to
  D^{(\ast)} \tau^- \bar{\nu}_\tau$ relative to $\bar{B} \to D^{(\ast)} \ell^-
  \bar{\nu}_\ell$ decays with hadronic tagging at Belle},'' {\em Phys. Rev.},
  vol.~D92, no.~7, p.~072014, 2015.

\bibitem{RD:abd}
A.~Abdesselam {\em et~al.}, ``{Measurement of the branching ratio of $\bar{B}^0
  \rightarrow D^{*+} \tau^- \bar{\nu}_{\tau}$ relative to $\bar{B}^0
  \rightarrow D^{*+} \ell^- \bar{\nu}_{\ell}$ decays with a semileptonic
  tagging method},'' in {\em {Proceedings, 51st Rencontres de Moriond on
  Electroweak Interactions and Unified Theories: La Thuile, Italy, March 12-19,
  2016}}, 2016.

\bibitem{Aaij:2017tyk}
R.~Aaij {\em et~al.}, ``{Measurement of the ratio of branching fractions
  $\mathcal{B}(B_c^+\,\to\,J/\psi\tau^+\nu_\tau)$/$\mathcal{B}(B_c^+\,\to\,J/\psi\mu^+\nu_\mu)$},''
  {\em Phys. Rev. Lett.}, vol.~120, no.~12, p.~121801, 2018.

\bibitem{Hiller:2003js}
G.~Hiller and F.~Kruger, ``{More model-independent analysis of $b \to s$
  processes},'' {\em Phys. Rev.}, vol.~D69, p.~074020, 2004.

\bibitem{Hiller:2014ula}
G.~Hiller and M.~Schmaltz, ``{Diagnosing lepton-nonuniversality in $b \to s
  \ell \ell$},'' {\em JHEP}, vol.~02, p.~055, 2015.

\bibitem{sdg}
S.~Descotes-Genon, L.~Hofer, J.~Matias, and J.~Virto, ``{Global analysis of
  $b\to s\ell\ell$ anomalies},'' {\em JHEP}, vol.~06, p.~092, 2016.

\bibitem{sdg2}
S.~Descotes-Genon, L.~Hofer, J.~Matias, and J.~Virto, ``{The $b \to sl^+
  l^{-1}$ Anomalies And Their Implications For New Physics},'' in {\em
  {Proceedings, 51st Rencontres de Moriond on Electroweak Interactions and
  Unified Theories: La Thuile, Italy, March 12-19, 2016}}, pp.~31--36, 2016.

\bibitem{bec}
D.~Bečirević, S.~Fajfer, and N.~Košnik, ``{Lepton flavor nonuniversality in
  $b\to s\ell^+\ell^-$ processes},'' {\em Phys. Rev.}, vol.~D92, no.~1,
  p.~014016, 2015.

\bibitem{Rk:eff}
W.~Altmannshofer and D.~M. Straub, ``{Implications of $b\to s$ measurements},''
  in {\em {Proceedings, 50th Rencontres de Moriond Electroweak Interactions and
  Unified Theories: La Thuile, Italy, March 14-21, 2015}}, pp.~333--338, 2015.

\bibitem{Rk:eff2}
G.~Hiller and M.~Schmaltz, ``{$R_K$ and future $b \to s \ell \ell$ physics
  beyond the standard model opportunities},'' {\em Phys. Rev.}, vol.~D90,
  p.~054014, 2014.

\bibitem{sim:eff}
B.~Bhattacharya, A.~Datta, D.~London, and S.~Shivashankara, ``{Simultaneous
  Explanation of the $R_K$ and $R(D^{(*)})$ Puzzles},'' {\em Phys. Lett.},
  vol.~B742, pp.~370--374, 2015.

\bibitem{Matias:2012xw}
J.~Matias, F.~Mescia, M.~Ramon, and J.~Virto, ``{Complete Anatomy of $\bar{B}_d
  -> \bar{K}^{* 0} (-> K \pi)l^+l^-$ and its angular distribution},'' {\em
  JHEP}, vol.~04, p.~104, 2012.

\bibitem{DescotesGenon:2012zf}
S.~Descotes-Genon, J.~Matias, M.~Ramon, and J.~Virto, ``{Implications from
  clean observables for the binned analysis of $B -> K*\mu^+\mu^-$ at large
  recoil},'' {\em JHEP}, vol.~01, p.~048, 2013.

\bibitem{Descotes-Genon:2013vna}
S.~Descotes-Genon, T.~Hurth, J.~Matias, and J.~Virto, ``{Optimizing the basis
  of $B\to K^*ll$ observables in the full kinematic range},'' {\em JHEP},
  vol.~05, p.~137, 2013.

\bibitem{Aaij:2015oid}
R.~Aaij {\em et~al.}, ``{Angular analysis of the $B^{0} \to K^{*0} \mu^{+}
  \mu^{-}$ decay using 3 fb$^{-1}$ of integrated luminosity},'' {\em JHEP},
  vol.~02, p.~104, 2016.

\bibitem{Aaij:2015esa}
R.~Aaij {\em et~al.}, ``{Angular analysis and differential branching fraction
  of the decay $B^0_s\to\phi\mu^+\mu^-$},'' {\em JHEP}, vol.~09, p.~179, 2015.

\bibitem{Wehle:2016yoi}
S.~Wehle {\em et~al.}, ``{Lepton-Flavor-Dependent Angular Analysis of $B\to
  K^\ast \ell^+\ell^-$},'' {\em Phys. Rev. Lett.}, vol.~118, no.~11, p.~111801,
  2017.

\bibitem{Abdesselam:2016llu}
A.~Abdesselam {\em et~al.}, ``{Angular analysis of $B^0 \to K^\ast(892)^0
  \ell^+ \ell^-$},'' in {\em {Proceedings, LHCSki 2016 - A First Discussion of
  13 TeV Results: Obergurgl, Austria, April 10-15, 2016}}, 2016.

\bibitem{Aaij:2014pli}
R.~Aaij {\em et~al.}, ``{Differential branching fractions and isospin
  asymmetries of $B \to K^{(*)} \mu^+ \mu^-$ decays},'' {\em JHEP}, vol.~06,
  p.~133, 2014.

\bibitem{Capdevila:2017bsm}
B.~Capdevila, A.~Crivellin, S.~Descotes-Genon, J.~Matias, and J.~Virto,
  ``{Patterns of New Physics in $b\to s\ell^+\ell^-$ transitions in the light
  of recent data},'' {\em JHEP}, vol.~01, p.~093, 2018.

\bibitem{Altmannshofer:2017yso}
W.~Altmannshofer, P.~Stangl, and D.~M. Straub, ``{Interpreting Hints for Lepton
  Flavor Universality Violation},'' {\em Phys. Rev.}, vol.~D96, no.~5,
  p.~055008, 2017.

\bibitem{Geng:2017svp}
L.-S. Geng, B.~Grinstein, S.~Jäger, J.~Martin~Camalich, X.-L. Ren, and R.-X.
  Shi, ``{Towards the discovery of new physics with lepton-universality ratios
  of $b\to s\ell\ell$ decays},'' {\em Phys. Rev.}, vol.~D96, no.~9, p.~093006,
  2017.

\bibitem{Ciuchini:2017mik}
M.~Ciuchini, A.~M. Coutinho, M.~Fedele, E.~Franco, A.~Paul, L.~Silvestrini, and
  M.~Valli, ``{On Flavourful Easter eggs for New Physics hunger and Lepton
  Flavour Universality violation},'' {\em Eur. Phys. J.}, vol.~C77, no.~10,
  p.~688, 2017.

\bibitem{Hurth:2017hxg}
T.~Hurth, F.~Mahmoudi, D.~Martinez~Santos, and S.~Neshatpour, ``{Lepton
  nonuniversality in exclusive $b{\rightarrow}s{\ell}{\ell}$ decays},'' {\em
  Phys. Rev.}, vol.~D96, no.~9, p.~095034, 2017.

\bibitem{Alok:2017jaf}
A.~K. Alok, D.~Kumar, J.~Kumar, and R.~Sharma, ``{Lepton flavor
  non-universality in the B-sector: a global analyses of various new physics
  models},'' 2017.

\bibitem{Alok:2017sui}
A.~K. Alok, B.~Bhattacharya, A.~Datta, D.~Kumar, J.~Kumar, and D.~London,
  ``{New Physics in $b \to s \mu^+ \mu^-$ after the Measurement of
  $R_{K^*}$},'' {\em Phys. Rev.}, vol.~D96, no.~9, p.~095009, 2017.

\bibitem{Choudhury:2017qyt}
D.~Choudhury, A.~Kundu, R.~Mandal, and R.~Sinha, ``{Minimal unified resolution
  to $R_{K^{(*)}}$ and $R(D^{(*)})$ anomalies with lepton mixing},'' {\em Phys.
  Rev. Lett.}, vol.~119, no.~15, p.~151801, 2017.

\bibitem{Buttazzo:2017ixm}
D.~Buttazzo, A.~Greljo, G.~Isidori, and D.~Marzocca, ``{B-physics anomalies: a
  guide to combined explanations},'' {\em JHEP}, vol.~11, p.~044, 2017.

\bibitem{lq1}
S.~Fajfer and N.~Košnik, ``{Vector leptoquark resolution of $R_K$ and
  $R_{D^{(*)}}$ puzzles},'' {\em Phys. Lett.}, vol.~B755, pp.~270--274, 2016.

\bibitem{lq2}
R.~Barbieri, G.~Isidori, A.~Pattori, and F.~Senia, ``{Anomalies in $B$-decays
  and $U(2)$ flavour symmetry},'' {\em Eur. Phys. J.}, vol.~C76, no.~2, p.~67,
  2016.

\bibitem{lq3}
M.~Bauer and M.~Neubert, ``{Minimal Leptoquark Explanation for the
  R$_{D^{(*)}}$ , R$_K$ , and $(g-2)_g$ Anomalies},'' {\em Phys. Rev. Lett.},
  vol.~116, no.~14, p.~141802, 2016.

\bibitem{lq4}
D.~Bečirević, N.~Košnik, O.~Sumensari, and R.~Zukanovich~Funchal,
  ``{Palatable Leptoquark Scenarios for Lepton Flavor Violation in Exclusive
  $b\to s\ell_1\ell_2$ modes},'' {\em JHEP}, vol.~11, p.~035, 2016.

\bibitem{lq5}
D.~Bečirević, S.~Fajfer, N.~Košnik, and O.~Sumensari, ``{Leptoquark model to
  explain the $B$-physics anomalies, $R_K$ and $R_D$},'' {\em Phys. Rev.},
  vol.~D94, no.~11, p.~115021, 2016.

\bibitem{lq6}
D.~Bečirević and O.~Sumensari, ``{A leptoquark model to accommodate
  $R_K^\mathrm{exp} < R_K^\mathrm{SM}$ and $R_{K^\ast}^\mathrm{exp} <
  R_{K^\ast}^\mathrm{SM}$},'' {\em JHEP}, vol.~08, p.~104, 2017.

\bibitem{DiLuzio:2017vat}
L.~Di~Luzio, A.~Greljo, and M.~Nardecchia, ``{Gauge leptoquark as the origin of
  B-physics anomalies},'' {\em Phys. Rev.}, vol.~D96, no.~11, p.~115011, 2017.

\bibitem{Calibbi:2017qbu}
L.~Calibbi, A.~Crivellin, and T.~Li, ``{A model of vector leptoquarks in view
  of the $B$-physics anomalies},'' 2017.

\bibitem{Crivellin:2015mga}
A.~Crivellin, G.~D'Ambrosio, and J.~Heeck, ``{Explaining $h\to\mu^\pm\tau^\mp$,
  $B\to K^* \mu^+\mu^-$ and $B\to K \mu^+\mu^-/B\to K e^+e^-$ in a
  two-Higgs-doublet model with gauged $L_\mu-L_\tau$},'' {\em Phys. Rev.
  Lett.}, vol.~114, p.~151801, 2015.

\bibitem{Crivellin:2015lwa}
A.~Crivellin, G.~D'Ambrosio, and J.~Heeck, ``{Addressing the LHC flavor
  anomalies with horizontal gauge symmetries},'' {\em Phys. Rev.}, vol.~D91,
  no.~7, p.~075006, 2015.

\bibitem{Crivellin:2015era}
A.~Crivellin, L.~Hofer, J.~Matias, U.~Nierste, S.~Pokorski, and J.~Rosiek,
  ``{Lepton-flavour violating $B$ decays in generic $Z'$ models},'' {\em Phys.
  Rev.}, vol.~D92, no.~5, p.~054013, 2015.

\bibitem{Greljo:2015mma}
A.~Greljo, G.~Isidori, and D.~Marzocca, ``{On the breaking of Lepton Flavor
  Universality in B decays},'' {\em JHEP}, vol.~07, p.~142, 2015.

\bibitem{Boucenna:2016wpr}
S.~M. Boucenna, A.~Celis, J.~Fuentes-Martin, A.~Vicente, and J.~Virto,
  ``{Non-abelian gauge extensions for B-decay anomalies},'' {\em Phys. Lett.},
  vol.~B760, pp.~214--219, 2016.

\bibitem{Pisano:1991ee}
F.~Pisano and V.~Pleitez, ``{An SU(3) x U(1) model for electroweak
  interactions},'' {\em Phys. Rev.}, vol.~D46, pp.~410--417, 1992.

\bibitem{Frampton:1992wt}
P.~H. Frampton, ``Chiral dilepton model and the flavor question,'' {\em Phys.
  Rev. Lett.}, vol.~69, pp.~2889--2891, Nov 1992.

\bibitem{Lee:1977qs}
B.~W. Lee and S.~Weinberg, ``Su(3) \ensuremath{\bigotimes} u(1) gauge theory of
  the weak and electromagnetic interactions,'' {\em Phys. Rev. Lett.}, vol.~38,
  pp.~1237--1240, May 1977.

\bibitem{Lee:1977tx}
B.~W. Lee and R.~E. Shrock, ``Su(3) \ensuremath{\bigotimes} u(1) gauge theory
  of weak and electromagnetic interactions,'' {\em Phys. Rev. D}, vol.~17,
  pp.~2410--2448, May 1978.

\bibitem{Buccella:1977gx}
F.~Buccella, M.~Lusignoli, and A.~Pugliese, ``Unified su(3)
  \ensuremath{\bigotimes} u(1) gauge theory with different muon and electron
  neutral currents,'' {\em Phys. Rev. Lett.}, vol.~40, pp.~1475--1478, Jun
  1978.

\bibitem{Buccella:1978nc}
F.~Buccella, L.~Caruso, and A.~Pugliese, ``The two neutral currents of the
  unified gauge group su(6),'' {\em Physics Letters B}, vol.~74, no.~4, pp.~357
  -- 360, 1978.

\bibitem{331:buras}
A.~J. Buras, F.~De~Fazio, J.~Girrbach, and M.~V. Carlucci, ``{The Anatomy of
  Quark Flavour Observables in 331 Models in the Flavour Precision Era},'' {\em
  JHEP}, vol.~02, p.~023, 2013.

\bibitem{331:burasmu}
A.~J. Buras, F.~De~Fazio, and J.~Girrbach, ``{331 models facing new $b \to
  s\mu^+ \mu^-$ data},'' {\em JHEP}, vol.~02, p.~112, 2014.

\bibitem{331:burasZZ}
A.~J. Buras, F.~De~Fazio, and J.~Girrbach-Noe, ``{$Z$-$Z'$ mixing and
  $Z$-mediated FCNCs in $SU(3)_{C} \times SU(3)_{L} \times U(1)_{X}$ models},''
  {\em JHEP}, vol.~08, p.~039, 2014.

\bibitem{331:anomaly}
R.~A. Diaz, R.~Martinez, and F.~Ochoa, ``{SU(3)(c) x SU(3)(L) x U(1)(X) models
  for beta arbitrary and families with mirror fermions},'' {\em Phys. Rev.},
  vol.~D72, p.~035018, 2005.

\bibitem{331:higgs}
R.~A. Diaz, R.~Martinez, and F.~Ochoa, ``{The Scalar sector of the SU(3)(c) x
  SU(3)(L) x U(1)(X) model},'' {\em Phys. Rev.}, vol.~D69, p.~095009, 2004.

\bibitem{ponce:331}
W.~A. Ponce, J.~B. Florez, and L.~A. Sanchez, ``{Analysis of SU(3)(c) x
  SU(3)(L) x U(1)(X) local gauge theory},'' {\em Int. J. Mod. Phys.}, vol.~A17,
  pp.~643--660, 2002.

\bibitem{sher:331}
D.~L. Anderson and M.~Sher, ``{3-3-1 models with unique lepton generations},''
  {\em Phys. Rev.}, vol.~D72, p.~095014, 2005.

\bibitem{Queiroz:2016gif}
F.~S. Queiroz, C.~Siqueira, and J.~W.~F. Valle, ``{Constraining Flavor Changing
  Interactions from LHC Run-2 Dilepton Bounds with Vector Mediators},'' {\em
  Phys. Lett.}, vol.~B763, pp.~269--274, 2016.

\bibitem{Alves:2016fqe}
A.~Alves, G.~Arcadi, P.~V. Dong, L.~Duarte, F.~S. Queiroz, and J.~W.~F. Valle,
  ``{Matter-parity as a residual gauge symmetry: Probing a theory of
  cosmological dark matter},'' {\em Phys. Lett.}, vol.~B772, pp.~825--831,
  2017.

\bibitem{331:CAB1}
J.~M. Cabarcas, J.~Duarte, and J.~A. Rodriguez, ``{Lepton Flavor Violation
  processes in 331 Models},'' {\em PoS}, vol.~HQL2012, p.~072, 2012.

\bibitem{331:CAB2}
J.~M. Cabarcas, J.~Duarte, and J.~A. Rodriguez, ``{Charged lepton mixing
  processes in 331 Models},'' {\em Int. J. Mod. Phys.}, vol.~A29, p.~1450015,
  2014.

\bibitem{Correia:2017vxa}
F.~C. Correia, ``{Fundamentals of the 3-3-1 Model with Heavy Leptons},'' {\em
  J. Phys.}, vol.~G45, no.~4, p.~043001, 2018.

\bibitem{Ng:1992st}
D.~Ng, ``{The Electroweak theory of SU(3) x U(1)},'' {\em Phys. Rev.},
  vol.~D49, pp.~4805--4811, 1994.

\bibitem{Grinstein:1987vj}
B.~Grinstein, R.~Springer, and M.~B. Wise, ``Effective hamiltonian for weak
  radiative b-meson decay,'' {\em Physics Letters B}, vol.~202, no.~1, pp.~138
  -- 144, 1988.

\bibitem{Buchalla:1995vs}
G.~Buchalla, A.~J. Buras, and M.~E. Lautenbacher, ``{Weak decays beyond leading
  logarithms},'' {\em Rev. Mod. Phys.}, vol.~68, pp.~1125--1144, 1996.

\bibitem{Glashow:2014iga}
S.~L. Glashow, D.~Guadagnoli, and K.~Lane, ``{Lepton Flavor Violation in $B$
  Decays?},'' {\em Phys. Rev. Lett.}, vol.~114, p.~091801, 2015.

\bibitem{Gauld:2013qba}
R.~Gauld, F.~Goertz, and U.~Haisch, ``{On minimal $Z'$ explanations of the
  $B\to K^*\mu^+\mu^-$ anomaly},'' {\em Phys. Rev.}, vol.~D89, p.~015005, 2014.

\bibitem{Gauld:2013qja}
R.~Gauld, F.~Goertz, and U.~Haisch, ``{An explicit Z'-boson explanation of the
  $B \to K^* \mu^+ \mu^-$ anomaly},'' {\em JHEP}, vol.~01, p.~069, 2014.

\bibitem{Lenz:2010gu}
A.~Lenz, U.~Nierste, J.~Charles, S.~Descotes-Genon, A.~Jantsch, C.~Kaufhold,
  H.~Lacker, S.~Monteil, V.~Niess, and S.~T'Jampens, ``{Anatomy of New Physics
  in $B - \bar{B}$ mixing},'' {\em Phys. Rev.}, vol.~D83, p.~036004, 2011.

\bibitem{Lenz:2012az}
A.~Lenz, U.~Nierste, J.~Charles, S.~Descotes-Genon, H.~Lacker, S.~Monteil,
  V.~Niess, and S.~T'Jampens, ``{Constraints on new physics in $B-\bar{B}$
  mixing in the light of recent LHCb data},'' {\em Phys. Rev.}, vol.~D86,
  p.~033008, 2012.

\bibitem{Charles:2013aka}
J.~Charles, S.~Descotes-Genon, Z.~Ligeti, S.~Monteil, M.~Papucci, and
  K.~Trabelsi, ``{Future sensitivity to new physics in $B_d, B_s$, and K
  mixings},'' {\em Phys. Rev.}, vol.~D89, no.~3, p.~033016, 2014.

\bibitem{Patrignani:2016xqp}
C.~Patrignani {\em et~al.}, ``{Review of Particle Physics},'' {\em Chin.
  Phys.}, vol.~C40, no.~10, p.~100001, 2016.

\bibitem{Schael:2013ita}
S.~Schael {\em et~al.}, ``{Electroweak Measurements in Electron-Positron
  Collisions at W-Boson-Pair Energies at LEP},'' {\em Phys. Rept.}, vol.~532,
  pp.~119--244, 2013.

\bibitem{ATLAS:2017wce}
T.~A. collaboration, ``{Search for new high-mass phenomena in the dilepton
  final state using 36.1 fb$^{-1}$ of proton-proton collision data at $\sqrt{s}
  =$ 13 TeV with the ATLAS detector},'' 2017.

\bibitem{Greljo:2017vvb}
A.~Greljo and D.~Marzocca, ``{High-$p_T$ dilepton tails and flavor physics},''
  {\em Eur. Phys. J.}, vol.~C77, no.~8, p.~548, 2017.

\bibitem{Grossman:1996qj}
Y.~Grossman, Z.~Ligeti, and E.~Nardi, ``{B ---> tau+ tau- (X) decays: First
  constraints and phenomenological implications},'' {\em Phys. Rev.}, vol.~D55,
  pp.~2768--2773, 1997.

\bibitem{Bobeth:2011st}
C.~Bobeth and U.~Haisch, ``{New Physics in $\Gamma_{12}^s$: ($\bar{s}
  b$)$(\bar{\tau} \tau)$ Operators},'' {\em Acta Phys. Polon.}, vol.~B44,
  pp.~127--176, 2013.

\bibitem{Alonso:2015sja}
R.~Alonso, B.~Grinstein, and J.~Martin~Camalich, ``{Lepton universality
  violation and lepton flavor conservation in $B$-meson decays},'' {\em JHEP},
  vol.~10, p.~184, 2015.

\bibitem{Kamenik:2017ghi}
J.~F. Kamenik, S.~Monteil, A.~Semkiv, and L.~V. Silva, ``{Lepton polarization
  asymmetries in rare semi-tauonic $ b \rightarrow s $ exclusive decays at
  FCC-$ee$},'' {\em Eur. Phys. J.}, vol.~C77, no.~10, p.~701, 2017.

\bibitem{Freytsis:2015qca}
M.~Freytsis, Z.~Ligeti, and J.~T. Ruderman, ``{Flavor models for $\bar{B} \to
  D^{(*)} \tau \bar{\nu}$},'' {\em Phys. Rev.}, vol.~D92, no.~5, p.~054018,
  2015.

\bibitem{Watanabe:2017mip}
R.~Watanabe, ``{New Physics effect on $B_c \to J/\psi \tau\bar\nu$ in relation
  to the $R_{D^{(*)}}$ anomaly},'' {\em Phys. Lett.}, vol.~B776, pp.~5--9,
  2018.

\bibitem{Bobeth:2007dw}
C.~Bobeth, G.~Hiller, and G.~Piranishvili, ``{Angular distributions of $\bar{B}
  \to \bar{K} \ell^+\ell^-$ decays},'' {\em JHEP}, vol.~12, p.~040, 2007.

\end{thebibliography}
\end{document}